\documentstyle[emulateapj,flushrt,epsf]{article}
\input psfig

\long\def\jumpover#1{{}}

\def \ngth{\negthinspace}
\def \ngth2{\negthinspace\negthinspace\negthinspace}

\def \th2{\thinspace\thinspace}

\def \Teff{{$T_{\rm ef\!f} $}}

\def \Mo{{$M_\odot $}}
\def \Lo{{$L_\odot $}}

\def \at{{\rm\char'100}}
\def \apriori{{\it a priori\ }}
\def \eg{{{\it e.g.},\ }}
\def \etal{{\it et al.\ }}

\def \ie{{{\it i.e.},\ }}

\def \viz{{\it viz.\ }}

\def \approxgt{\,\raise2pt \hbox{$>$}\kern-8pt\lower2.pt\hbox{$\sim$}\,}
\def \approxlt{\,\raise2pt \hbox{$<$}\kern-8pt\lower2.pt\hbox{$\sim$}\,}
\def \th{\thinspace}

\def \Apriori{{\it A priori\ }}
\def \apriori{{\it a priori\ }}

\begin{document}

\submitted{ASTROPHYSICAL JOURNAL, revised May 2004}
\title{Evidence for Low-Dimensional Chaos in Semiregular Variable Stars}
\author{ J. Robert Buchler$^{1}$\
         Zolt\'an Koll\'ath$^{2}$\&
         Robert R. Cadmus, Jr.$^{3}$
        }
\begin{abstract}
                 An analysis of the photometric observations of the light
curves of the five large amplitude, irregularly pulsating stars R~UMi, RS~Cyg,
V~CVn, UX~Dra and SX~Her is presented.  First, multi-periodicity is eliminated
for these pulsations, \ie they are not caused by the excitation of a small
number of pulsation modes with constant amplitudes.  Next, on the basis of
energetics we also eliminate stochasticity as a cause, leaving low dimensional
chaos as the only alternative.  We then use a global flow reconstruction
technique in an attempt to extract quantitative information from the light
curves, and to uncover common physical features in this class of irregular
variable stars that straddle the RV~Tau to the Mira variables.  Evidence is
presented that the pulsational behavior of R~UMi, RS~Cyg, V~CVn and UX~Dra
takes place in a 4-dimensional dynamical phase space, suggesting that two
vibrational modes are involved in the pulsation.  A linear stability analysis
of the fixed points of the maps further indicates the existence of a two-mode
resonance, similar to the one we had uncovered earlier in R~Sct:~~The irregular
pulsations are the result of a continual energy exchange between two strongly
nonadiabatic modes, a lower frequency pulsation mode and an overtone that are
in a close 2:1 resonance.  The evidence is particularly convincing for R~UMi,
RS~Cyg and V~CVn, but much weaker for UX~Dra.  In contrast, the pulsations of
SX~Her appear to be more complex and may require a 6D space.  
\end{abstract}


\date{May 2004}



 {\bigskip
        {\footnotesize
 \noindent $^1$Physics Department, University of Florida, Gainesville, FL, USA;
 buchler\at phys.ufl.edu \\
 \noindent $^2$Konkoly Observatory, Budapest, HUNGARY; kollath\at konkoly.hu\\
 \noindent $^2$Department of Physics, Grinnell College, Grinnell, IA 50112,
 USA}
 }


\section{INTRODUCTION}

The five stars considered here are pulsating variables and show the sorts of
light curve irregularities that are generally associated with classification as
semiregular variables.  The measurements section of the SIMBAD database lists V
CVn, R UMi, and RS Cyg as SRa stars, UX Dra as SRb, and SX Her as SRd, although
conflicting classifications exist.  RS Cyg and UX Dra are also classified as
carbon stars.  SX Her is earlier in spectral type (K2) than the others (late M
or C).

V CVn has been reported to have two closely spaced periods (Loeser et al. 1986,
Kiss et al. 1999, Kiss et al.  2000) and exhibits time-dependent polarization
(Boyle et al. 1986, Magalh�es et al. 1986).  SX Her is a metal-poor star
(Preston and Wallerstein 1963) that shows no signs of an infrared excess
(Carney et al.  2003).  Its period is apparently changing slowly (Percy and
Kolin 2000).

In \S3 we argue on physical grounds that the irregular pulsations of these
stars cannot be multi-periodic as has been suggested (Kiss \etal 1999), nor can
they be of a stochastic nature as proposed in Konig \etal (1999).  This means
that, by default, they therefore must be chaotic.  If the light curve is indeed
unsteady because of a chatic dynamics, then it is not astonishing that an
interpretation within a traditional periodic framework would lead to a spurious
conclusion that there are closely spaced peaks or that the period is changing
(Buchler, J. R. \& Koll\'ath, Z., 2001).

In \S4 we apply a global flow reconstruction (\eg Serre \etal 1996) to the
data.  This tool has been specifically designed for the study of chaotic
signals which have only a few degrees of freedom (low dimensional chaos).  We
show that this reconstruction can capture the pulsation dynamics of these
stars.  The fact that the reconstruction allows us to generate synthetic LCs
with properties similar to the observed LCs thus provides additional, {\it a
posteriori} evidence for low dimensional chaos.

The global flow reconstruction also yields quantitative information about the
underlying dynamics, such as its dimension.  We can use this information to
uncover the physical nature of the irregular LCs of these stars.  In
particular, it allows us to address the question of whether the same resonant
mechanism is operative here as in the case of the RV~Tau-type star R~Scuti
where the large amplitude irregular pulsations were shown to be a consequence
of the resonant excitation of an overtone with close to double the frequency of
the basic pulsation mode under highly nonadiabatic conditions (Buchler \etal
1995, 1996).  The fundamental mode is self-excited and wants to grow, whereas
the entrained, resonant overtone is stable and wants to decay, from which a
chaotic motion of alternating growth and decay ensues, a scenario known in
nonlinear dynamics under the name of Shilnikov (\eg Berg\'e \etal 1986).



 \begin{figure*}
 \centerline{\psfig{figure=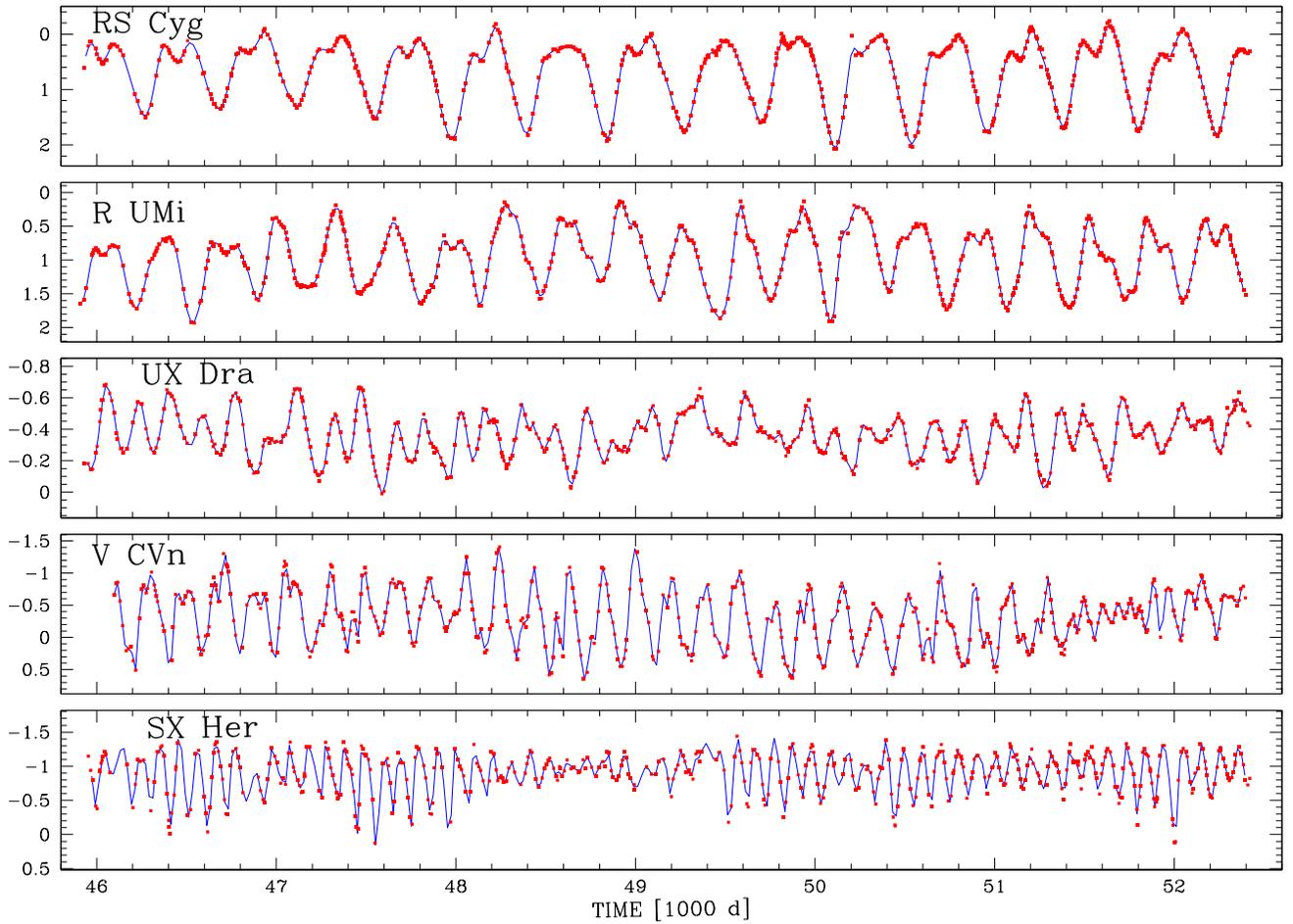,width=17.3cm}}
  \parbox[b]{18cm}{
 \caption[] {\small {\bf Light curves (mag).}
   Dots: Observed LCs  (Cadmus);
  solid lines: fits to LCs, binned and averaged over 10\th d,
  and smoothed with $\sigma_s=0.01$ (see text);
  Time = Julian date 2400000 + [1000\th days].
   }
 \vskip -5pt
 \label{cad}
 }
\end{figure*}


\section{Observations}

Our analysis is based on high quality V band light curve (LC) data from the
observational program at Grinnell College that span close to 18 years.  The
observations were made with the 0.61 m telescope at the Grant O.  Gale
Observatory using a conventional photoelectric photometer.  The data were
reduced differentially relative to the following comparison stars (in
parentheses after the associated variable star): R UMi (HD 150747), RS Cyg (HD
192422), V CVn (HD 116531), UX Dra (HD 178089), SX Her (HD 145226).  The
comparison stars were monitored relative to additional check stars and found to
exhibit no significant variability.  The reduction included corrections for
atmospheric extinction and transformation to the UBV system using B band data
that are not presented here.  The precision of the data is determined primarily
by atmospheric conditions but can be estimated to be about 0.01 magnitude from
the repeated measurements that make up a single data point as well as from
related data.

Because continuity of the data is important for the sort of analysis presented
here, efforts were made to essentially eliminate the gaps between observing
seasons by taking some data (a small fraction of the total) at larger airmasses
and in brighter twilight than is desirable for photometric observations.  The
quality of the data obtained in twilight was improved by modeling the nonlinear
time variation in the sky brightness.  Although the quality of the measurements
obtained under these conditions is a bit worse than that for data taken under
more favorable conditions, it is still good enough for the purposes of this
paper and is far preferable to gaps in the data.  The data are available from
RRC.

The observed LC data are displayed as dots, jointly with the smoothed fits, in
Fig. 1.  The sampling is irregular with an average spacing of ~ 10 d.  The
light curves of R UMi, V CVn, and UX Dra exhibit sizeable vertical shifts that
could be due to dust formation during the pulsation but, as the next section
suggests, they could also be part of the intrinsic pulsation dynamics of the
stars.  The LC of SX Her shows large fluctuations in overall amplitude.  RS Cyg
displays distinctive dips at the maxima of its LC.


\section{THE NATURE OF THE PULSATIONS}

\subsection{Time-Frequency Analysis}

\begin{figure*}
\centerline{\vbox{\epsfxsize=8.7cm\epsfbox{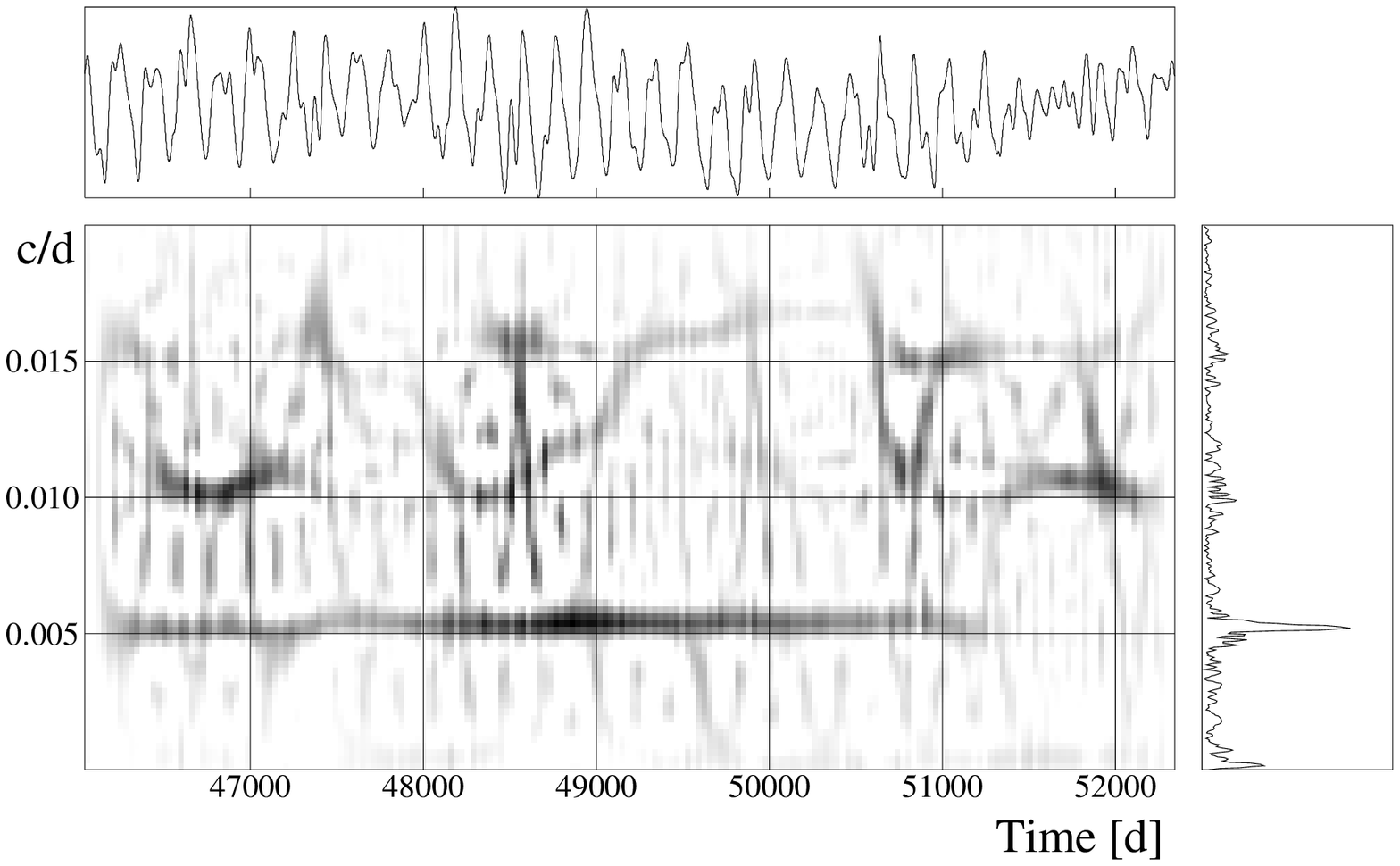}}\hskip 5pt
              \vbox{\epsfxsize=8.7cm\epsfbox{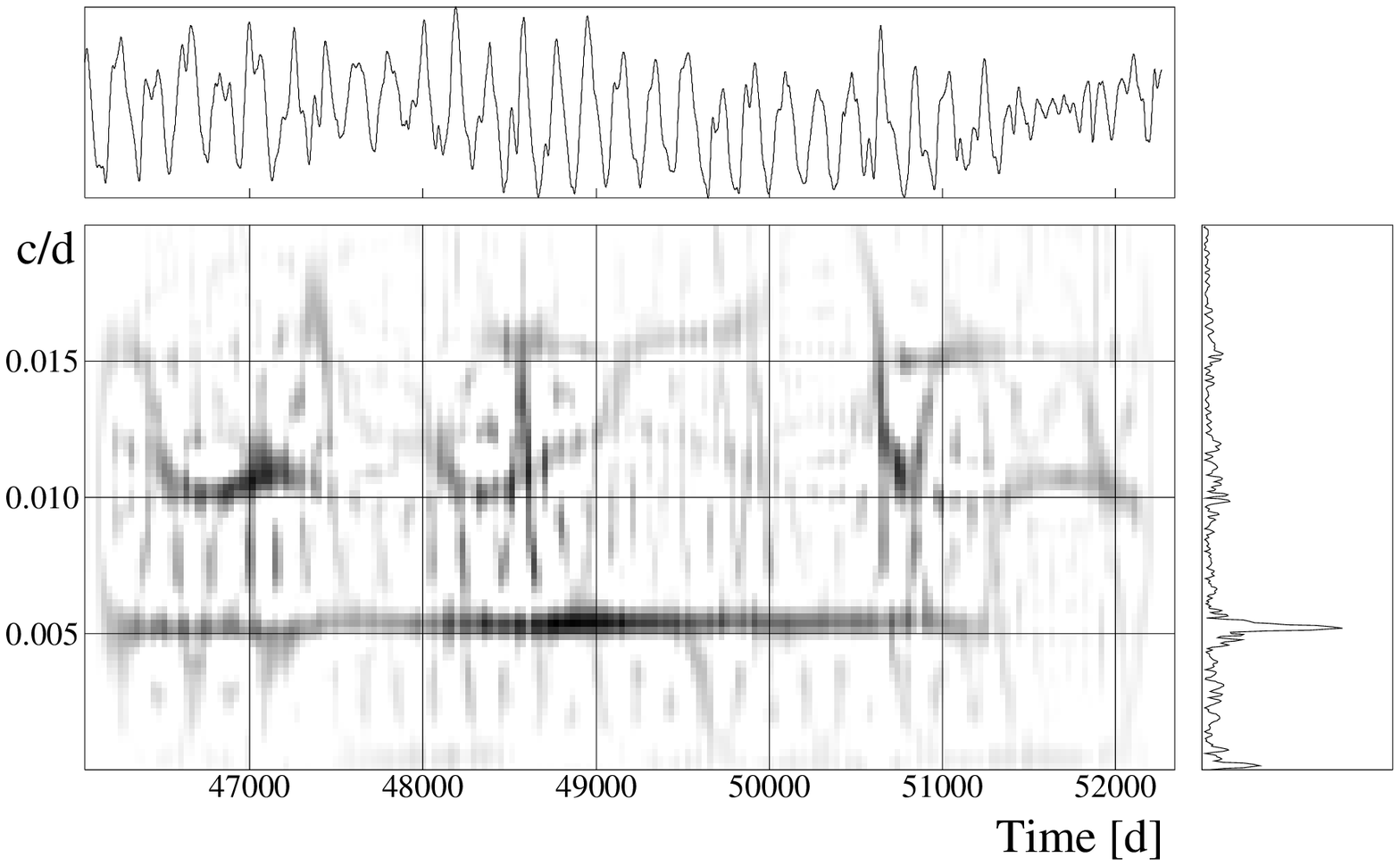}}}\caption[]{\small {\bf V~CVn:}
{\bf Left Figure:} Cadmus photoelectric LC;
{\bf Right Figure:} combined visual amateur astronomer data.
In each figure: {\sl Top}: smoothed LCs,
{\sl right}: amplitude Fourier spectrum,
{\sl center}: time-frequency plot.
   }
 \label{vcvnzk}
\end{figure*}

Figure~\ref{vcvnzk}, we present a comparison of the photoelectric Cadmus data
and the combined visual amateur astronomer data (AAVSO, AFOEV, BAAVS, VSJSO) of
V~CVn for approximately the same time-intervals.  These smoothed time-series
therefore have a different zero points and different amplitudes, and they also
have quite a different quality.  (The smoothings that we apply to these data
will be discussed below.)

We display the smoothed LCs (top), the amplitude Fourier spectra (FS)
(right) and the time-frequency (TF) plots (Koll\'ath \& Buchler 1996), also
known as spectrograms.  For the latter we have used the Zhao-Atlas-Marks (1990)
kernel that belongs to a class of TFs known under the name of Reduced
Interference Distributions (RIDs).  These are designed to give much sharper
images than the older transforms, such as G\'abor, for example (e.g. Cohen
1994).  In the TF plots we have enhanced the power at the harmonics to the same
level as the fundamental to make it visible ($A(f) = A_{TF}(f) S(f)$, where
$S=1$ for $f< 1.1 f_0$, where $f_0$ is the frequency of the fundamental peak,
$S=5$, for example, for $f>1.9 f_0$, and a linear increase in between).  A
comparison of the TF plots with the FS (on the right) clearly shows the size of
this enhancement $S$.

Despite the fact that the Cadmus and amateur LCs are appreciably different at
times, the TF plots are remarkably similar, showing their insensitivity to
observational errors and to the smoothing procedure.

\begin{figure*}
\centerline{\vbox{\epsfxsize=8.7cm\epsfbox{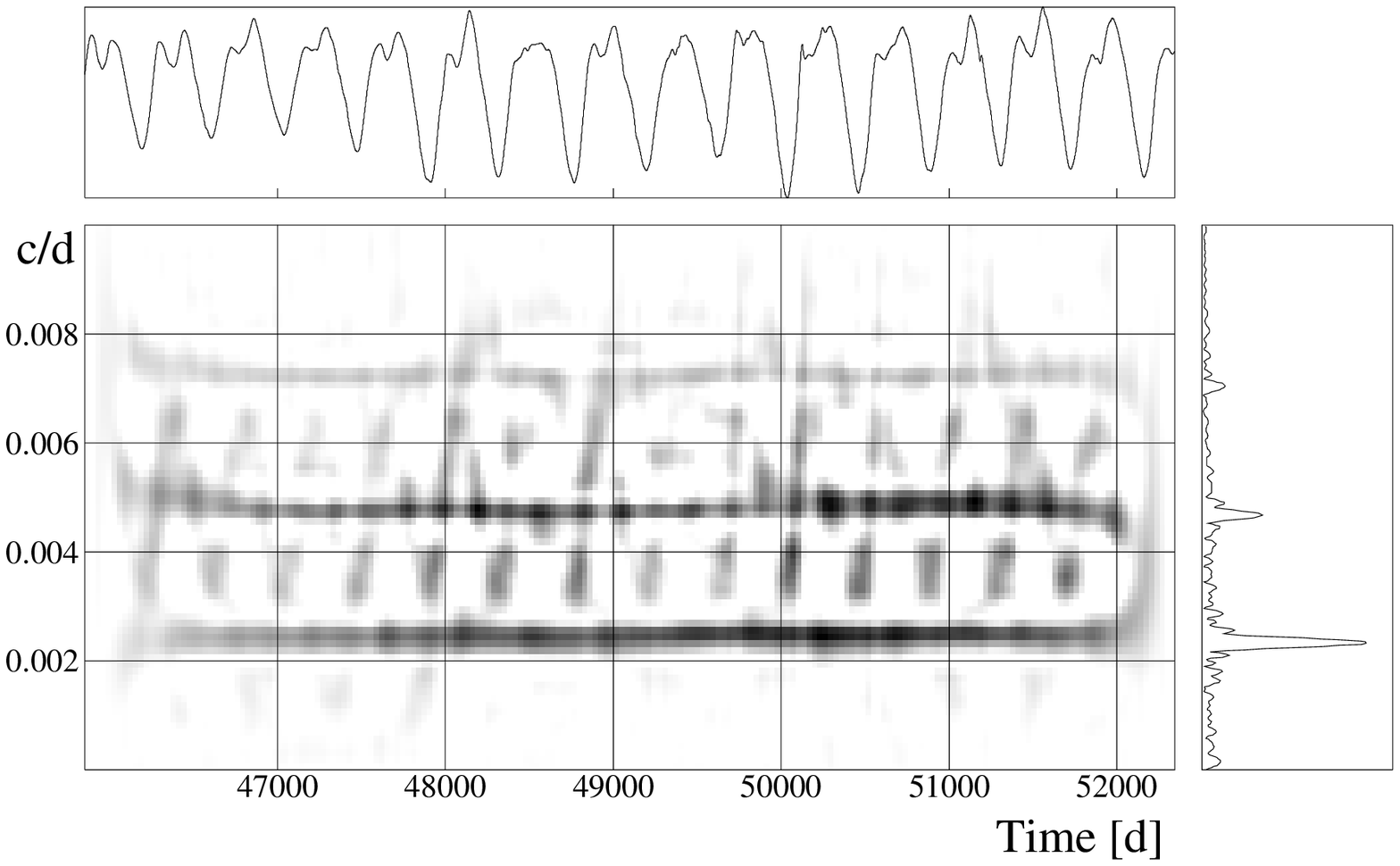}}\hskip 5pt
              \vbox{\epsfxsize=8.7cm\epsfbox{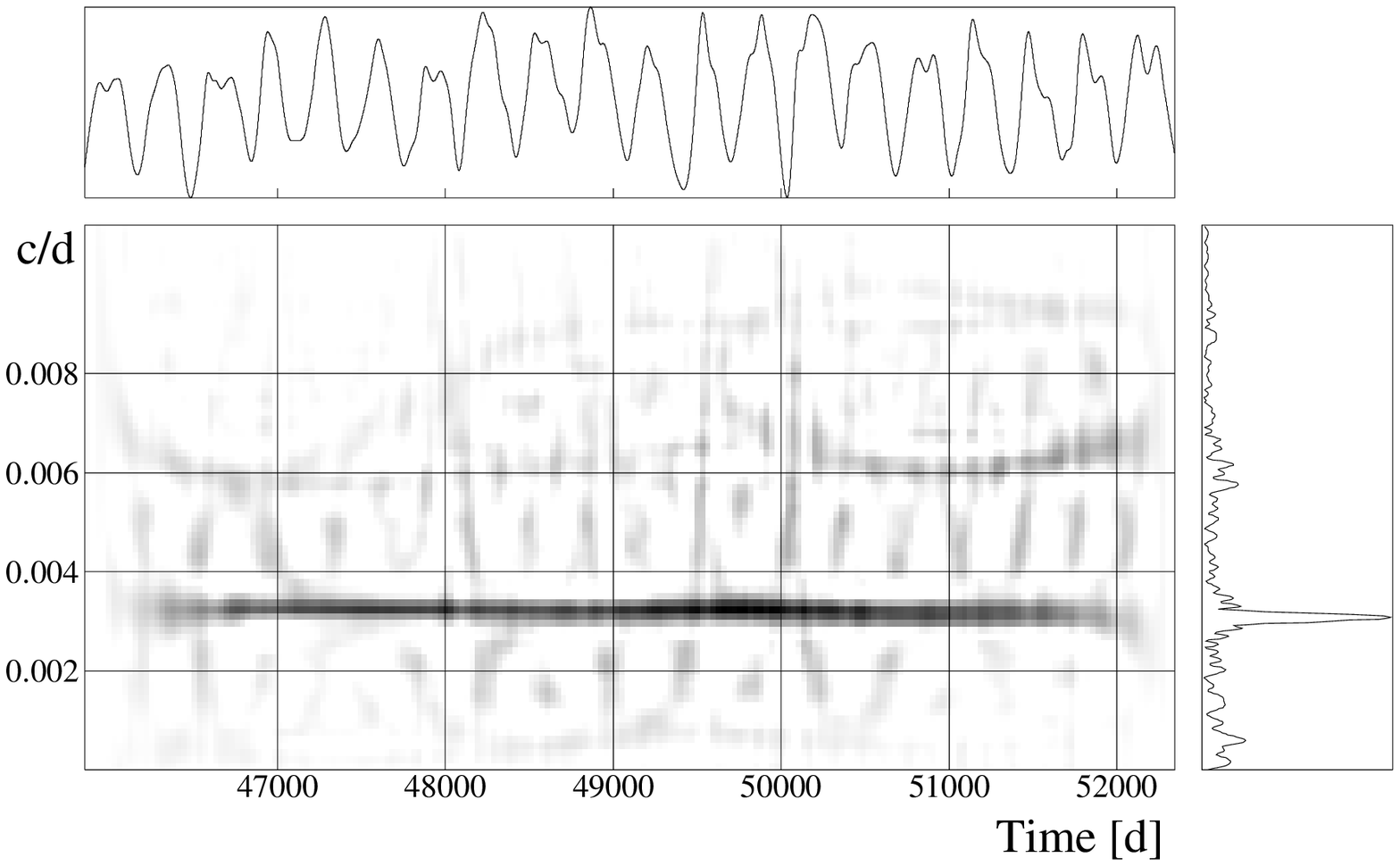 }}}
\caption[]{\small {\bf Left Figure:} {\bf RS~Cyg:} and
{\bf Right Figure:} {\bf R~UMi}.
{\sl Top}: smoothed LCs,
{\sl right}: amplitude Fourier spectra,
{\sl center}: time-frequency plots.
   }
 \label{rscygzk}
 \end{figure*}

\begin{figure*}
\centerline{\vbox{\epsfxsize=8.7cm\epsfbox{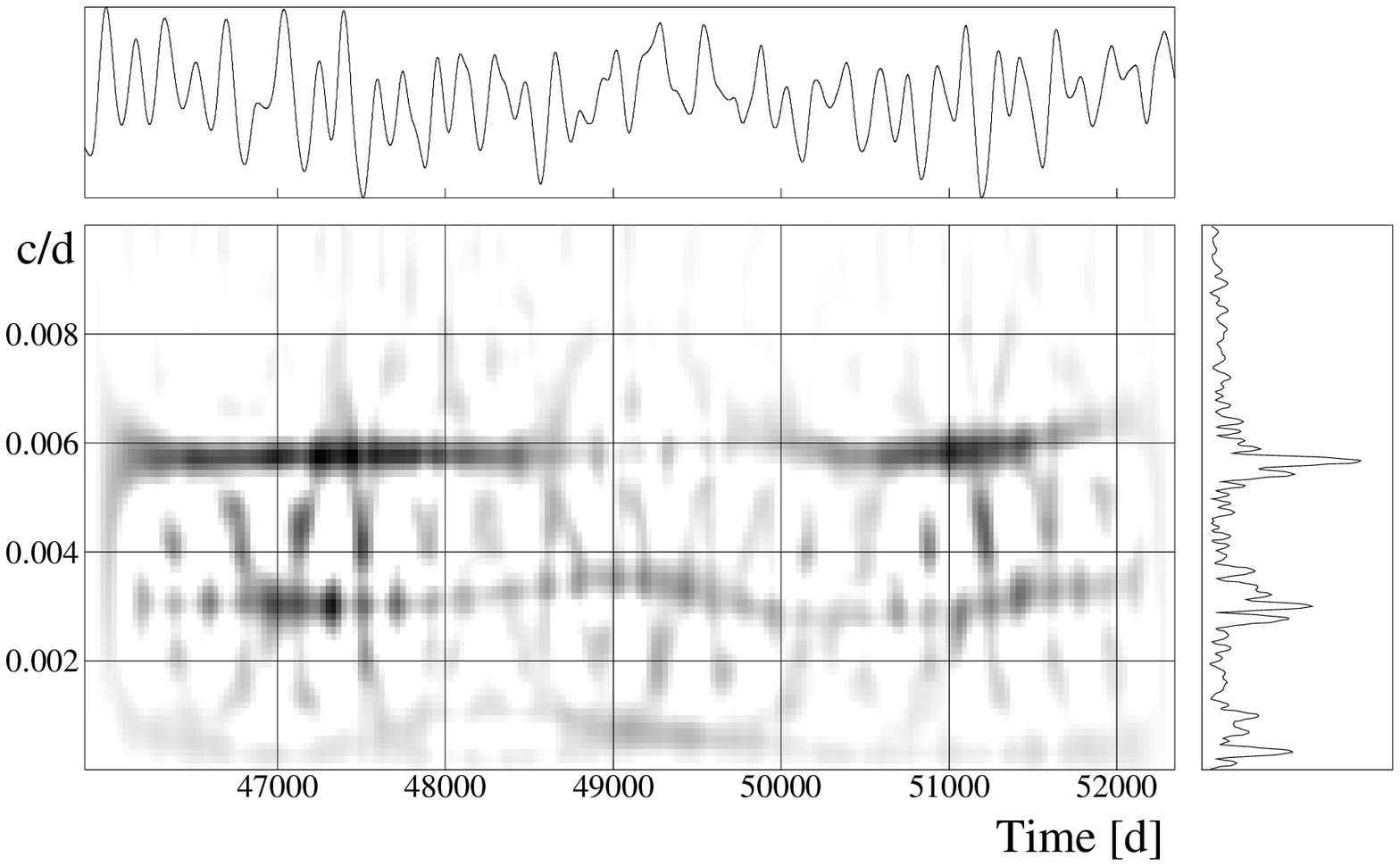}}\hskip 5pt
              \vbox{\epsfxsize=8.7cm\epsfbox{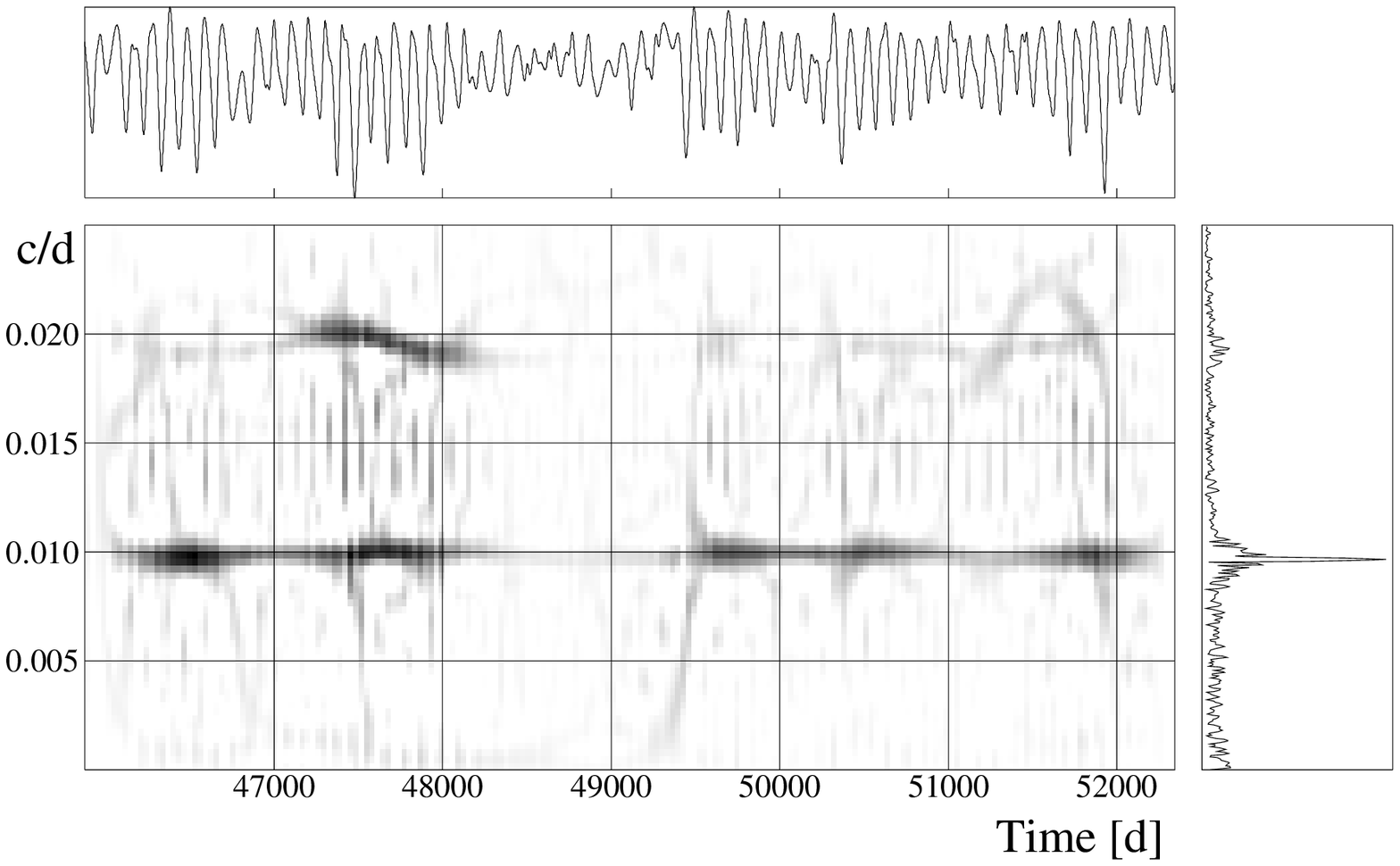}}}
\caption[]{\small {\bf Left Figure:} {\bf UX Dra}
and {\bf Right Figure:} {\bf SX~Her}.
{\sl Top}:  smoothed LCs,
{\sl right}:  amplitude Fourier spectra,
{\sl center}: time-frequency plota.
   }
 \label{sxherzk}
 \end{figure*}

In Figs.~\ref{vcvnzk} through \ref{sxherzk} we display the TFs of RS~Cyg,
R~UMi, UX~Dra and SX~Her.  The TFs of all five stars show some amplitude
(power) at twice the fundamental peak, and both RS~Cyg and V~CVn also have
appreciable amplitude at $3f_0$.  From the very appearance of the observed LCs
one finds it hardly astonishing that for all five stars the instantaneous
amplitudes in the dominant peaks waver in time, but more interestingly the
instantaneous frequencies change as well.  In addition, the first harmonic
frequency peak does not change synchronously with that of the fundamental
frequency.  This asynchrony between the fundamental and the harmonic peaks
rules out that all of the harmonic structure is due to the anharmonicity of the
motion.  Furthermore it cannot be explained by dust or spots, by binarity, by
evolution or by stochasticity, even though all of these effects may be present
and affect the LCs.

We shall first argue that the pulsations of these stars are neither periodic
nor multi-periodic (see also Buchler \& Koll\'ath 2003).  Then we will give
physical, energetic arguments that also eliminate a stochastic nature for these
pulsations.  This then, by default, leaves a low dimensional chaotic nature as
the only alternative.


 \vskip 20pt

\subsection{Multi-periodicity?}

Because the 'multi-periodic' label is sometimes used with different
connotations, to be specific, we mean that the signal is such that (a)
the FS contains a few discrete, basic frequencies, and all the other
peaks correspond to linear combinations of these basic frequencies
that arise because of nonlinear couplings, and (b) the amplitudes and
frequencies of all these peaks are constant, or, at most, vary very slowly
in time.  The beat Cepheids and RR Lyrae are examples of this type of systems.

Multi-periodicity would thus imply that the major frequency peaks $f_k$ in the
Fourier spectra are real and should correspond to actual pulsation frequencies
or linear combinations thereof.  Using these frequency peaks one can then make
a multi-periodic fit $s(t) = \sum A_k sin(2\pi f_k t +\phi_k)$ to the
observational data.  For the stars that we are considering we find that we need
from 50 to 100 frequencies for a decent fit.  This however poses a physical
problem, namely the origin of these frequencies.  Indeed, stellar modelling
shows that there simply do not exist enough radial modes in the frequency range
of interest for these types of star.  This remains true even if allowance is
made for nonradial modes with $\ell$ low enough (say $\ell = 3$) to be
observable photometrically and assuming that the latter can indeed be excited
to such large amplitudes.  Furthermore, typically only a few modes are found to
be linearly unstable, and one would have to explain why so many other stable
modes get entrained to such high amplitudes.  We thus conclude that these LCs
are not generated by a large set of excited pulsation modes.

There is still the possibility of a multi-periodic, but slowly evolving star.
The structure of the peaks should thus changes slowly as well, but for our five
stars the timespan of the observational data is too short compared to the
evolutionary timescale.  We note however, that for the RV~Tau type star R~Sct
(Koll\'ath 1990), the frequencies in the FS appear randomly variable from one
section of the data to another which thus also eliminates an explanation in
terms of slowly evolving multi-periodic stars.

\subsection{Stochasticity}

A recent paper (Konig \etal 1999) claim to explain the irregular behavior of
R~Sct as the superposition of two stochastically excited damped linear
oscillators.  Typically stochastic excitation of vibrationally stable modes,
such as the solar oscillations, leads to much smaller amplitudes.  It seems
more likely that pulsations as large as those observed in R~Sct are due to a
self-excited mode, such as in the Cepheid and RR Lyrae variables.  In fact this
brings us to our main rejection of stochastic excitation for R~Sct.  While such
a stochastic 'explanation' is perhaps correct from a mathematical point of
view, it is not meaningful on physical grounds because no mechanism is proposed
that could excite damped modes to such large pulsation amplitudes (up to
factors of 40 in the LC of R~Sct, for example).

In order to be more quantitative, let us consider a 0.7\Mo\ model with
L=1000\Lo\ and \Teff=5300K (and X=0.726, Z=0.004).  In our turbulent convective
hydro-code (\eg Koll\'ath \etal 2002) this model is unstable in the fundamental
mode and leads to chaotic pulsations of alternating large and small amplitudes.
The average pulsational kinetic energies in such typical cycles are
$\langle$KE$\rangle$ = 0.31$\times 10^{42}$ and 0.23$\times 10^{42}$ ergs,
respectively.  The average turbulent energies are $\langle$TE$\rangle$ =
0.15$\times 10^{42}$ and 0.12$\times 10^{42}$ ergs, somewhat smaller, but of
the same order within the accuracy of mixing-length models.  The alleged
stochastic excitation of Konig \etal (1999) can only be associated with the
convective motions.  Our numbers show that even if {\sl all} available
turbulent energy were converted into pulsational kinetic energy, there is
barely enough energy.  The pulsation is closely associated with a single (most
likely radial) pulsation mode, or perhaps at most two of them, thus has a very
directed form of kinetic energy.  The amount of energy transferred
'stochastically' from the kinetic energy associated with the largely disordered
turbulent convection to these modes depends on the overlap integrals between
the spatial distribution of the turbulent energy (\eg Eqs. 6 and 7 of Buchler,
Goupil \& Kov\'acs 1993).  This overlap must be extremely small because of the
very different nature of the two types of motion.  Furthermore the power in the
Fourier spectrum of turbulent convection at the pulsation frequency is very
limited.  This rules out stochasticity as the cause of the observed irregular
behavior.

However, the amount of energy flowing through the pulsating envelope, \viz the
product of the luminosity and the 'period' $\sim$6.1$\times 10^{42}$ ergs, is
almost 100 times the pulsation kinetic energy.  This highly nonadiabatic
situation certainly allows the observed changes in pulsational kinetic energy.
A perhaps better estimate of the efficiency of conversion of heat into
pulsation energy, or the percentage change of pulsation kinetic energy per
cycle, is given by the linear growth-rates of the modes (\eg $\kappa_0 \times
P_0 \sim 1.25$ for the stellar model above).  This is precisely what a chaotic
dynamics is about: The nonlinear interaction of a small number of strongly
nonadiabatic (dissipative) pulsation modes.

We conclude that it is therefore ultimately physics that rules out
stochasticity for these stars.  This is not to say, however, that superposed on
the low dimensional radial chaotic dynamics, there cannot be stochastic
motions, which furthermore are nonradial.

\subsection{Other Reasons For Low Dimensional Chaos}

There are additional reasons for believing that the pulsations of these five
stars are chaotic and have a low dimension: \th First, numerical hydrodynamic
simulations, albeit of W~Virginis type stellar models, uncovered the chaotic
nature of the pulsations (Buchler \& Kov\'acs 1987, Aikawa 1987, Buchler,
Goupil \& Kov\'acs 1987, Kov\'acs \& Buchler 1988, Letellier \etal 1996).
Second, the analysis of the LC of the RV~Tau-type star R~Scuti clearly
demonstrated the presence of a low dimensional chaotic pulsation dynamics
(Buchler, Serre, Koll\'ath \& Mattei 1995, Buchler, Koll\'ath, Serre \& Mattei
1996).  More recent work by other authors gives corroborating evidence for low
dimensional chaos in large amplitude variabel stars (Kiss \& Szatm\'ary 2002,
Ambika \etal 2003).

\vskip 5pt

Linear analyses are very useful for multi-periodic signals, but it is well
known, at least outside astronomy, that for irregular, intrinsically nonlinear
signals they yield limited information about the dynamics (\eg Weigend \&
Gershenfeld 1994, Abarbanel \etal 1993).  As we have already mentioned, a
phenomenologically motivated multiperiodic fit will always be 'successful' as
an interpolation, but will give us not much information about the stellar
pulsation aside from one or perhaps several average 'periods'.  For the study
such chaotic signals and for understanding the underlying dynamics a different
approach is required.

\vskip 5pt

A popular test for nonlinear behavior is the construction of first
return maps (\eg Ott 1993).  For a dynamics with a fractal
attractor with fractal dimension $\approxgt$2, such as the
R\"ossler band, the first return maps show a tight parabolic
structure, but for our stars they turn out to be essentially
scatter diagrams.  This negative result implies that, if this
attractor is indeed fractal, the fractal dimension of the dynamics
substantially exceeds a value of 2, and could even be greater than
3.  We recall that for the star R~Sct it was found that $d_L\sim
3.1$ (Buchler \etal 1996).


 \begin{figure*}
\centerline{\psfig{figure=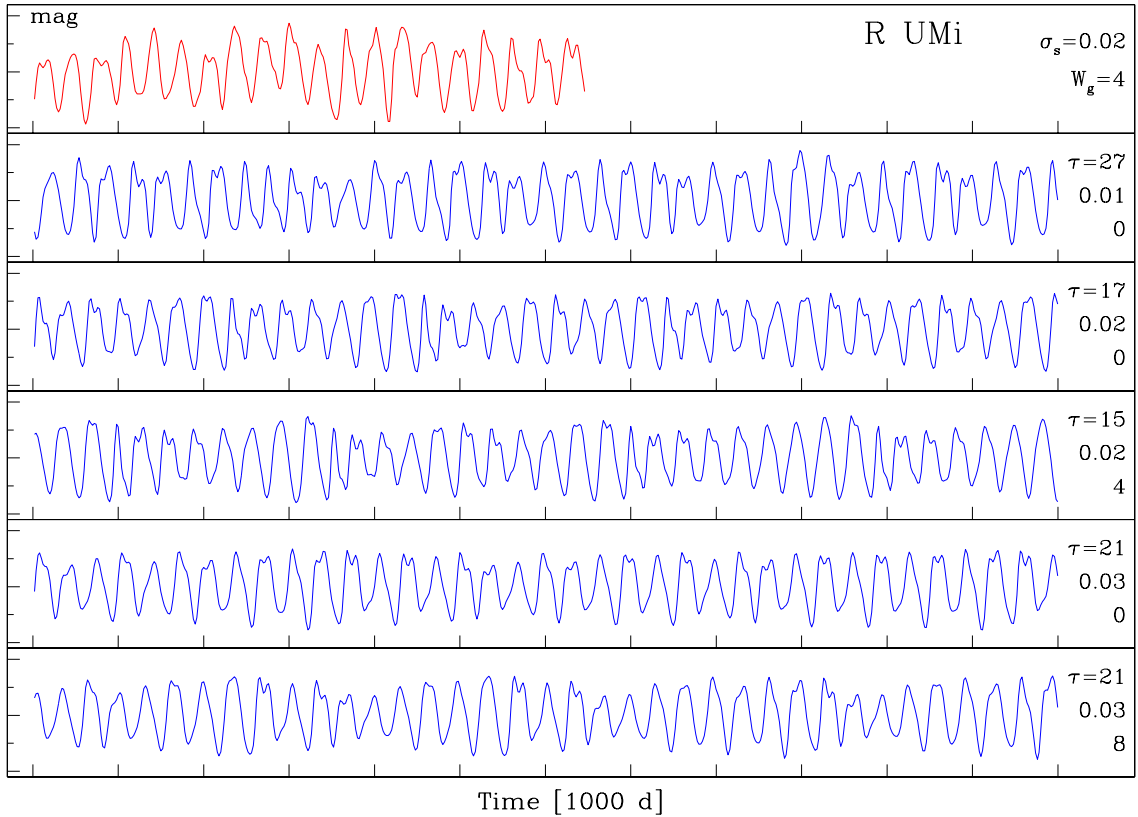,width=15.cm}}
\caption[]{\small {\bf R~UMi}, top: smoothed LC (mag);
below:
sections of synthetic LCs with various seeds and delays
$\tau$ for
maps constructed from the smoothed magnitude data;
($\tau$,$\sigma_s$,$W_g$) are indicated on the right
($W_g=0$ stands for no Gaussian filter);
   }
 \label{rumisyn}
 \end{figure*}


\begin{table*}
  \caption{
  Linear stability properties
  of the fixed points of the maps ${\mathbf \cal M}$ for a
  selected subset of synthetic LCs with smoothings
  $\sigma_s$, $W_g$, delays $\tau$: Given are the
  the locations of the main fixed points  $M_{fp}$, the growth rates $\kappa$,
  the frequencies   $f$ and the ratios $f_2/f_1$.
  All frequencies and growth rates are in  $[d^{-1}]$.
  }
  \begin{center}
 \begin{tabular}{llrrrccccr}
    \hline
    \hline
    \noalign{\smallskip}
   {\bf R~UMi}&  $\sigma_s$ & $W_g$  & $\tau$ & $M_{fp}$ &
        $\kappa_1$ & $f_1$ & $\kappa_2$ & $f_2$ &  $f_2/f_1$ \\
    \noalign{\smallskip}
    \hline
    \noalign{\smallskip}
    \noalign{\smallskip}
&0.01 & 0 & 23
  & 1.064
  &  2.093E--03 &  3.543E--03
  & --3.275E--03 &  7.340E--03
  &  2.07 \\
&0.01 & 0 & 25
  & 1.072
  &   2.442E--03 &  3.504E--03
  &  --3.635E--03 &  7.254E--03
  &  2.07 \\
&0.01 & 8 & 19
  & 1.063
  &   1.508E--03 &  3.532E--03
  &  --2.712E--03 &  7.764E--03
  &   2.20 \\
&0.02 & 0 & 17
  & 1.081
  &   1.296E--03 &  3.511E--03
  &  --2.112E--03 &  7.898E--03
  &   2.25 \\
&0.02 & 4 & 15
  & 1.083
  &   1.319E--03 &  3.523E--03
  &  --2.146E--03 &  8.039E--03
  &   2.28 \\
&0.02 & 4 & 17
  & 1.087
  &   1.516E--03 &  3.539E--03
  &  --2.188E--03 &  7.890E--03
  &   2.23 \\
&0.03 & 0 & 21
   &1.096
  &   1.233E--03 &  3.554E--03
  &  --2.696E--03 &  6.759E--03
  &   1.90 \\
&0.03 & 8 & 21
  & 1.091
  &   1.420E--03 &  3.547E--03
  &  --3.231E--03 &  6.554E--03
  &   1.85   \\
    \noalign{\smallskip}
    \hline
    \hline
    \noalign{\smallskip}
   {\bf RS Cyg:}& $\sigma_s$ & $W_g$  & $\tau$ & $M_{fp}$ &
        $\kappa_1$ & $f_1$ & $\kappa_2$ & $f_2$ &  $f_2/f_1$ \\
    \noalign{\smallskip}
    \noalign{\smallskip}
    \hline
    \hline
    \noalign{\smallskip}
    \noalign{\smallskip}
&0.025 & 1 & 40 & 0.631
  & 3.368E--03 &  2.793E--03
  & +1.981E--03 &  5.248E--03
  & 1.88 \\
&0.025 & 7 & 40 & 0.622
  & 3.149E--03 &  2.793E--03
  & +1.830E--03 &  5.365E--03
  & 1.92 \\
&0.03 & 4 & 35 & 0.581
  &   3.539E--03  & 2.701E--03
  &  --5.122E--04  & 5.438E--03
  &  2.01  \\
&0.04 & 4 & 33 &  0.489
  &   -1.142E-03  & 6.245E-03
  &    2.398E-03  & 2.724E-03
  &   2.29  \\
&0.04 & 4 & 40 &  0.500
  &   1.258E--03  & 2.708E--03
  &  --9.522E--04  & 6.444E--03
  &   2.38  \\
&0.04 & 5  & 34 & 0.491
  &   2.214E--03  & 2.724E--03
  &  --1.062E--03  & 6.290E--03
  &   2.31 \\
    \noalign{\smallskip}
    \hline
    \hline
    \noalign{\smallskip}
   {\bf V CVn:}&   $\sigma_s$ & $W_g$  & $\tau$ & $M_{fp}$ &
        $\kappa_1$ & $f_1$ & $\kappa_2$ & $f_2$ &  $f_2/f_1$ \\
    \noalign{\smallskip}
    \hline
    \hline
    \noalign{\smallskip}
    \noalign{\smallskip}
&0.01 & 0 & 22
& --0.279
& --4.665E--03  & 5.102E--03
&  4.098E--03  & 1.119E--02
& 2.19 \\
&0.01 & 6 & 22
& --0.278
& --4.514E--03 &  5.087E--03
& 4.005E--03 &  1.105E--02
&2.17 \\
&0.01 & 6 & 23
& --0.263
& --4.933E--03 &  5.065E--03
&  4.108E--03  & 1.047E--02
&2.07 \\
&0.01 & 12 & 21
& --0.279
& --4.665E--03 &  5.102E--03
&  4.098E--03  & 1.119E--02
&2.19 \\
&0.02 & 0 & 22
& --0.279
& --4.672E--03 &  5.076E--03
&  4.694E--03  & 1.108E--02
&2.18 \\
&0.02 & 6 & 21
& --0.287
& --4.448E--03 &  5.124E--03
&  4.309E--03  & 1.155E--02
&2.25 \\
&0.03 & 0 &21
& --0.292
& --4.615E--03 &  5.115E--03
&  4.999E--03 &  1.152E--02
&2.25 \\
    \noalign{\smallskip}
    \hline
    \hline
    \noalign{\smallskip}
   {\bf UX Dra:}&   $\sigma_s$ & $W_g$  & $\tau$ & $M_{fp}$ &
        $\kappa_1$ & $f_1$ & $\kappa_2$ & $f_2$ &  $f_2/f_1$ \\
    \noalign{\smallskip}
    \hline
    \hline
    \noalign{\smallskip}
    \noalign{\smallskip}
&0.018 &0 & 6
& --0.39
& --1.236E--02 &  3.963E--03
&   8.458E--03 &  5.975E--03
& 1.51 \\
&0.016 &0 & 5
& --0.43
& --9.825E--03  & 3.478E--03
&   5.684E--03  & 7.398E--03
& 2.13  \\
&0.017& 0 & 6
& --0.41
& --1.155E--02 &  3.720E--03
&   7.904E--03  & 6.423E--03
& 1.73 \\
&0.018 &0 & 6
& --0.39
& --1.236E--02 &  3.963E--03
&   8.458E--03  & 5.975E--03
& 1.51 \\
&0.019 &0 & 5
& --0.39
& --1.522E--02 &  3.633E--03
&   1.058E--02  & 5.548E--03
& 1.53 \\
    \noalign{\smallskip}
    \hline
    \hline
  \end{tabular}
  \end{center}
\label{tab}
\end{table*}


\section{GLOBAL FLOW RECONSTRUCTION}

Over the last decade we have developed a special tool for the analysis of
irregular time-series, namely the {\sl global flow reconstruction technique}
(reconstruction technique for short).  Our primary motivation has not been
phenomenological, but rather has aimed for an understanding of the physical
mechanism behind the irregular LCs of variable stars on the one hand,
and for deriving qualitative and quantitative properties of this dynamics
(Serre \etal 1996; reviewed in Buchler 1997, Buchler \& Koll\'ath 2001; a
similar approach was followed in other research areas, \eg Gouesbet \etal 1997,
Hegger \etal 1999).


 \begin{figure*}
  \centerline{\vbox{\epsfxsize=8.1cm\epsfbox{
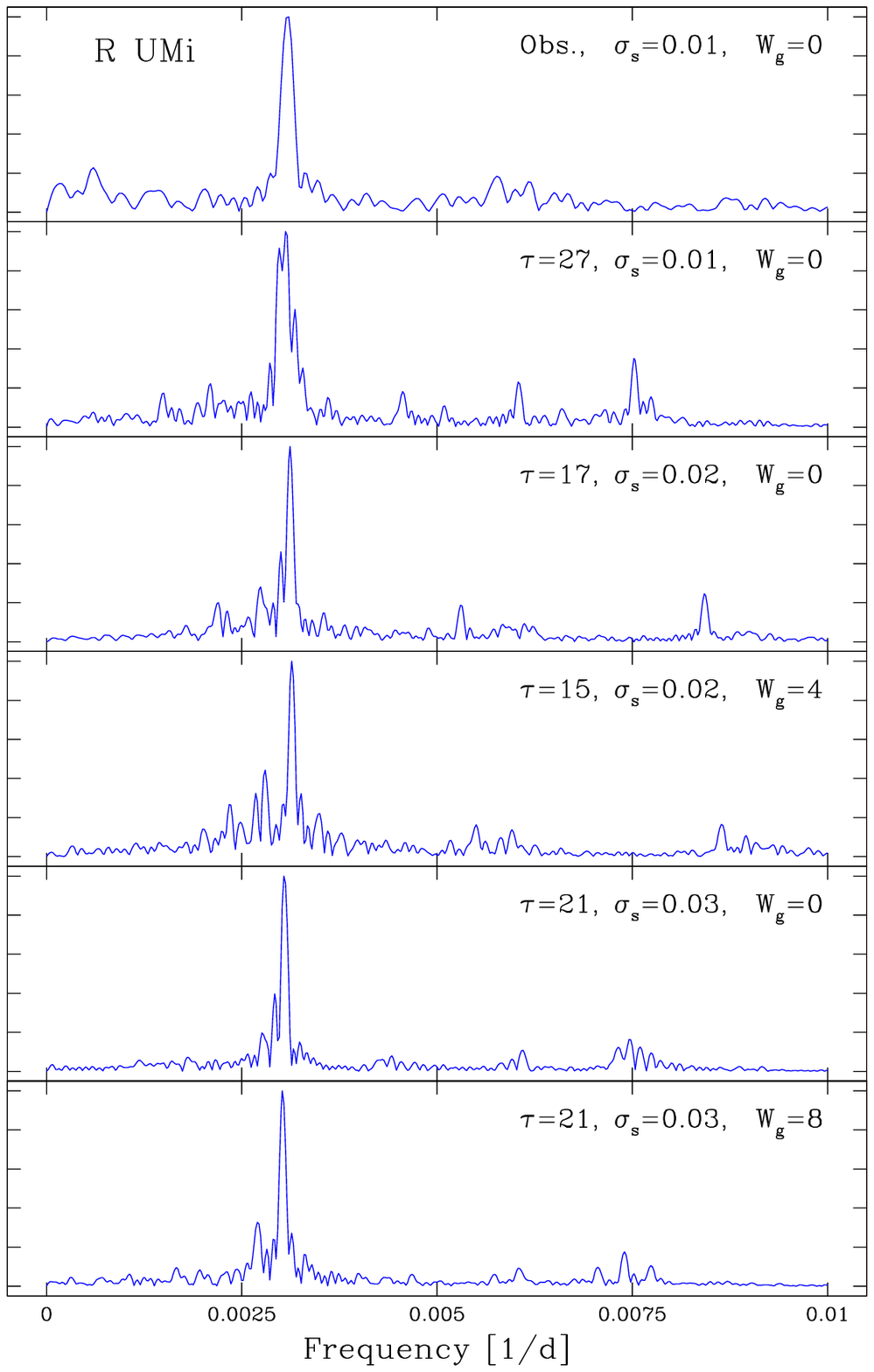
}}\hskip 5pt
              \vbox{\epsfxsize=6.55cm\epsfbox{
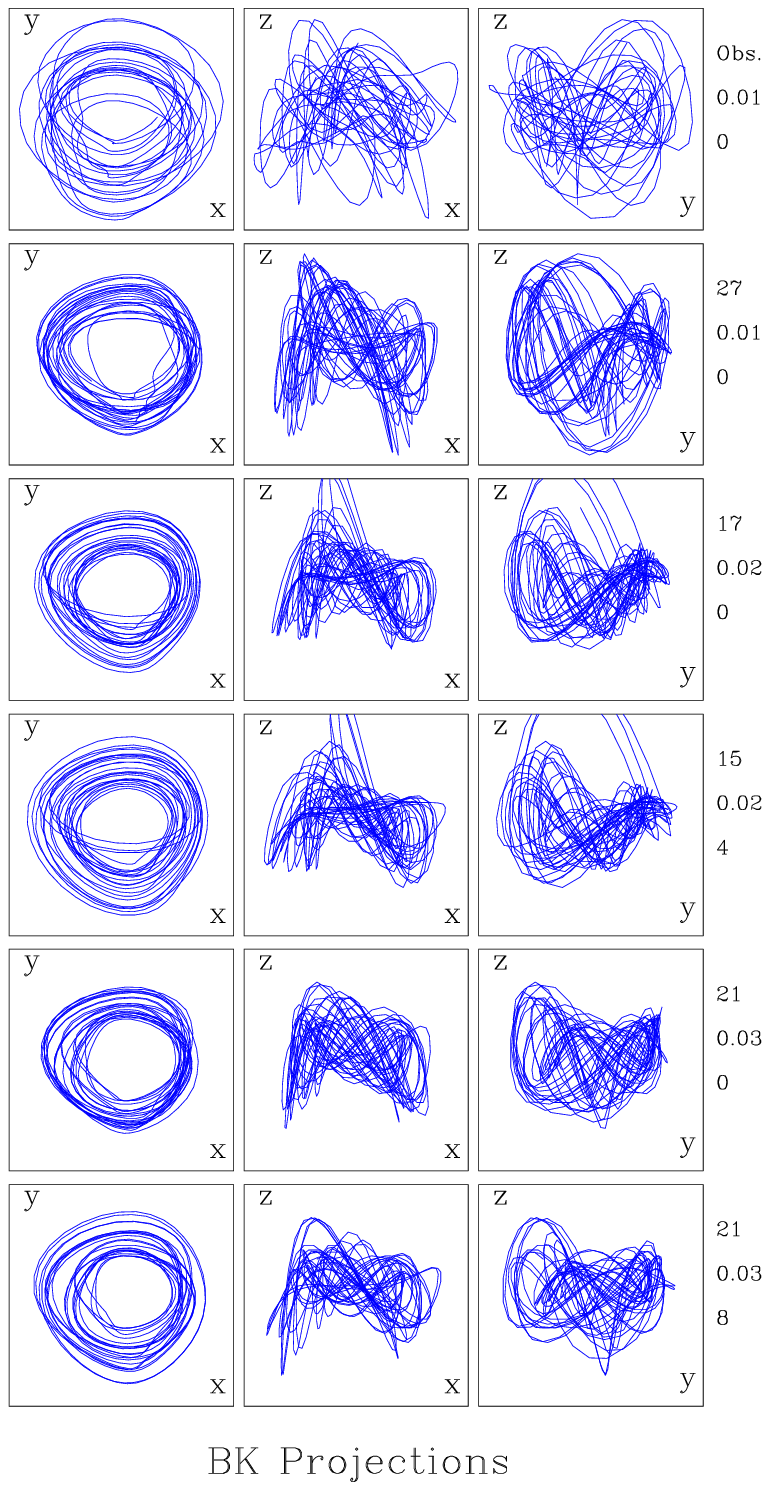
}}}
\caption[]
  {\small {\bf R UMi:}
   LEFT: Amplitude FS
   top: of the smoothed LC,
   below: of synthetic LCs.
 RIGHT:  corresponding Broomhead-King (BK) projections.
   }
 \label{rumifs}
 \end{figure*}

Only two very simple working assumptions underlie the reconstruction technique:
(1) {\sl the LC is produced by the (deterministic) nonlinear
interaction of a small number of pulsation modes}, (2) {\sl the system is
autonomous}, \ie explicit time-dependence, such as evolution over the span of
the data, can be ignored.

\Apriori it may appear astonishing that one should be able to derive basic
physical properties of the star from the mere knowledge of a {\sl scalar}
time-series, such as the LC.  Yet, in the case of the star R~Scuti it
was possible to demonstrate that the irregular pulsations are the result of the
nonlinear interaction between a self-excited pulsation mode that entrains a
vibrationally stable mode through an approximate 2:1 resonance (Buchler \etal
1995, 1996).

The technique introduces a (Euclidean) {\sl reconstruction space} of dimension
$d_R$, in which one uses the observed time-series
$s_n = s(t_n)$ to construct successive
position vectors
$${\bf~X_n}~=~(s_n,~s_{n-\tau},~s_{n-2\tau},~\ldots,~s_{n-(d_R-1)\tau}).$$

The quantity $\tau$ is called the delay.
Assumption (1) allows us to use the observed time-series to construct the
best mathematical relationship ${\bf \cal M}$, (called a map or a flow in the
case of a differential description), that evolves the system that produces the
LC from one time-step to the next:
$\quad {\bf X_{n+1}} = {\bf \cal M}({\bf X_n}).\quad $
A dimension $d_R$ that is large enough to capture the dynamics of
the pulsation is called an {\sl embedding dimension}.  Our goal is
to find the {\sl minimum embedding dimension} $d_e$.  This
dimension $d_e$ is not known \apriori and it is one of the
fundamental quantities that the analysis aims to determine.  This
dimension is important because it is closely related to the
dimension $d$ of the actual 'physical' phase space of the
dynamics.  For example, for a single oscillatory mode, $d$ would
be equal to 2, for 2 coupled oscillatory modes $d=4$, and the
involvement of a secular mode would add 1 to the dimension.
Practice has shown that a multivariate polynomial functional form
is suitable for the map ${\mathbf \cal M}$ and that polynomes up
to order of 4 are generally sufficient; the coefficients of these
polynomials are computed with a least-squares fit to the observed
time-series (Serre \etal 1996).  The reconstruction technique also
introduces a {\sl delay} $\tau$ which we consider a free
parameter.

The reconstruction requires a time series with equal time intervals.  Our data
preparation is done in several steps:~ (a) the data are binned and averaged
over a time interval $\Delta t_{av}$.  The typical time interval for the Cadmus
data has a very broad peak around 10\th d.  In the following we have chosen
values for $\Delta t_{av}$ = 10\th d, except for UX~Dra where we chose 1\th d.~
(b) The data are next smoothed with a cubic spline (Reinsch 1967) with a
smoothing $\sigma_s$ which results in an equally spaced time series, with time
intervals chosen to be 1\th d.  (c) Finally a moving Gaussian filter of width
$W_g$ is applied to the splined time series.

Results of a preliminary analysis of these stars were reported in Buchler,
Koll\'ath \& Cadmus 2001.  Here we perform a much broader analysis; in
particular our search covers a wider range of smoothings, $\sigma_s$ and
$W_g$.  As a result we obtain firmer conclusions than were obtained in that
paper.


 \begin{figure*}
\centerline{\psfig{figure=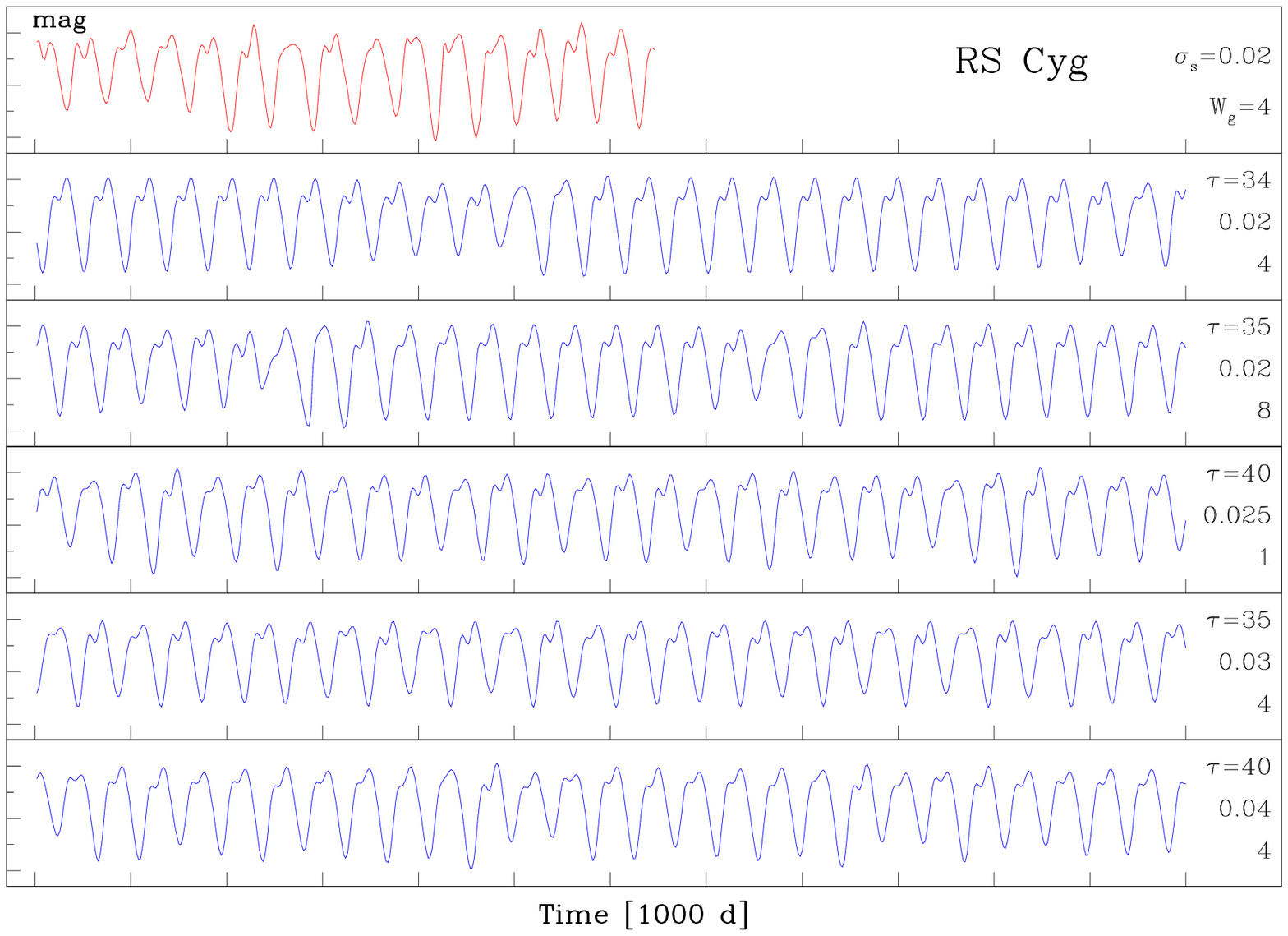,width=15.cm}}
\caption[]{\small {\bf RS~Cyg},
top: smoothed LC (mag);
below:
sections of synthetic LCs with various seeds and delays
$\tau$ for
maps constructed from the smoothed magnitude data;
($\tau$,$\sigma_s$,$W_g$) are indicated on the right;
   }
 \label{rscygsyn}
 \end{figure*}


 \begin{figure*}
 \centerline{\psfig{figure=
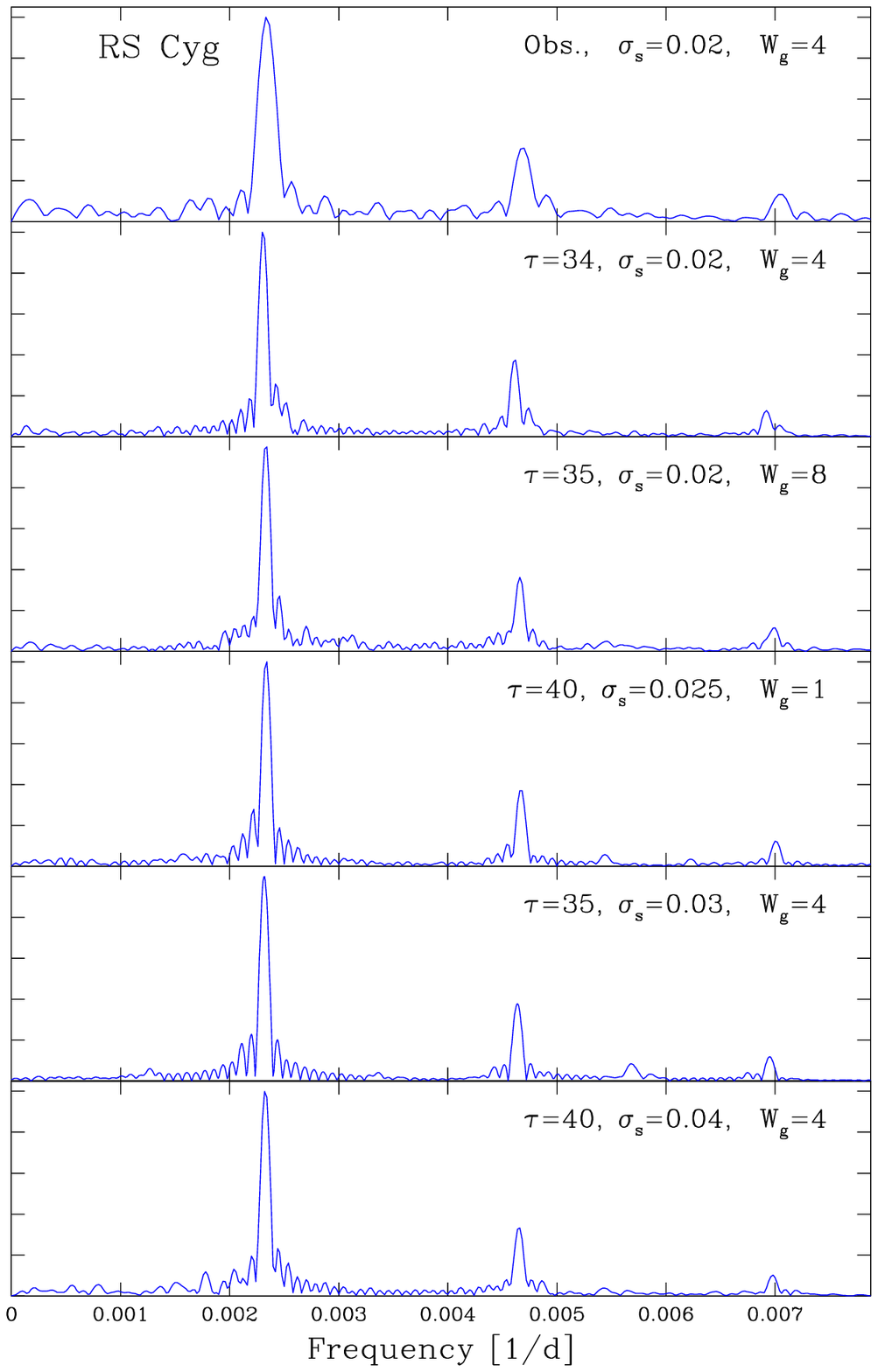
,width=8.1truecm}
             \psfig{figure=
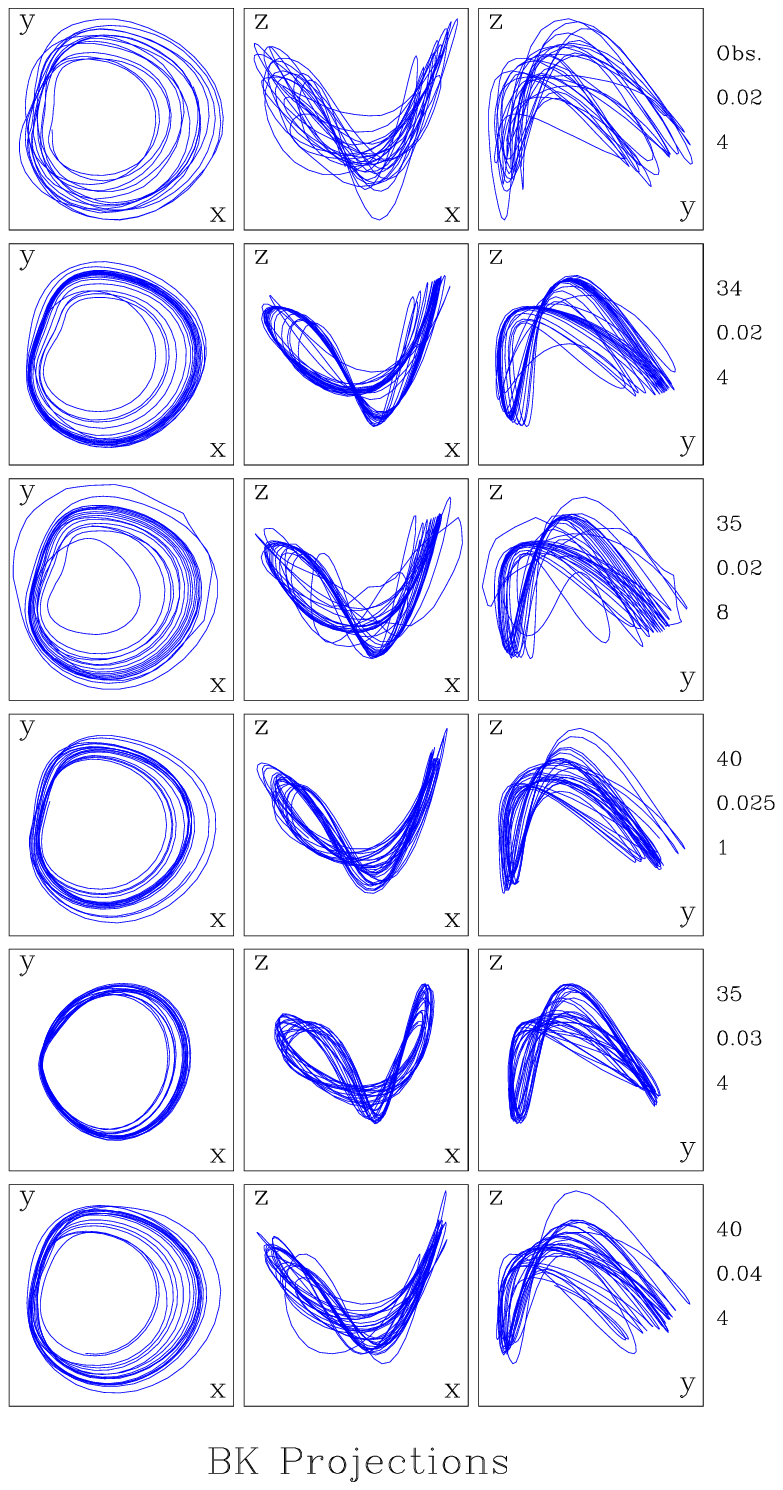
,width=6.62truecm} }
\caption[]{\small {\bf RS~Cyg}:
LEFT: Amplitude FS
top: of the smoothed LC,
below: of synthetic LCs.
 RIGHT:  corresponding Broomhead-King (BK) projections.
   }
 \label{rscygfs}
 \end{figure*}


\vskip -1.5cm
\subsection{\bf R~UMi}

In this section we apply the reconstruction to the observed LC of
R~UMi, and concurrently we discuss some of the technical details of the
reconstruction.

In Figure~\ref{rumisyn} on top we display a fit to the R~UMi data, with the
smoothing parameters $\sigma_s=0.02$ and $W_g=4$.  Two observational points, at
JD 46039.5095 and 46613.8169, give an unnatural looking wiggle in the smoothed
data (\eg in Fig.~1 of Buchler \etal 2001), and have therefore been omitted in
the reconstructions.  We have verified though that the reconstructions with all
the observational points give essentially the same results, indicating a
relative insensitivity of the R~UMi reconstructions to individual points.

In practice, the reconstructions proceed as follows: from a smoothed light
curve (with given $\sigma_s$ and $W_g$) a map ${\mathbf\cal M}$ is constructed
for a given dimension $d_R$ and a given value of delay $\tau$.  This map
${\mathbf\cal M}$ is then iterated from a seed value (chosen along the observed
time-series) to generate what we call a {\it synthetic} LC.  Actually, the map
${\mathbf\cal M}$ generates $d_R$ scalar sequences, namely the components of
${\bf X_n}$, shifted with respect to each other by $\tau$ days, but are
essentially identical when superposed.  This can be considered an {\it a
posteriori} confirmation that the reconstruction is successful.  For R~UMi as
for the other stars we have chosen 15000~d long iterations of which we
typically discard the first three thousand days, because they may be transients
in the evolution toward the chaotic attractor.

Because the data preparation obviously will have an effect on the reconstructed
attractors, we therefore perform them for a range of smoothing and filtering
values $\sigma_s$ and $W_g$.  Ideally, we would also expect the results of the
reconstructions to be independent of these smoothing parameters in some range
of values (For a discussion of these issues we refer to Serre \etal 1996).  In
our surveys we perform the reconstructions with delays typically ranging from
$\tau= 4$ to 40~d. In a robust reconstruction the results should come out
similar for some range of neighboring $\tau$ values.

It is obvious that for a successful reconstruction the data have to sample
sufficiently well the dominant features of the dynamics in phase space.  When
the LC is very short -- there are only 20 pulsation cycles for R~UMi -- the
iteration of the constructed maps is not always stable and the synthetic LCs
blow up. There can be several causes for such behavior, \eg 'boundary crises'
(\eg Ott 1993).  In practice, we iterate a given map successively with
different seeds until we obtain a synthetic LC of sufficient length and
quality.

How do we judge the quality of the reconstruction? \th\th There is no point in
comparing chaotic signals point by point, but rather they need to be compared
in a global or statistical sense.  However, we are seriously hampered by the
shortness of both the observed and synthetic LCs.  We are therefore forced to
rely on more subjective criteria.

Our first acceptance criterion relies on {\sl a visual comparison} of the
observed and synthetic LCs.  When we attempt reconstructions for R~UMi in 3D
(\ie with $d_R=3$), we find that the synthetic LCs bear no resemblance to the
observed LC for any $\sigma_s$ or $W_g$.  It is therefore immediately obvious
that $d_e>3$.

In contrast, the 4D reconstructions show a relatively large insensitivity to
the smoothing.  We have made our survey with $\sigma_s=0.01$, 0.02, to 0.05 and
$W_g=0$, 4 and 8 ($W_g$\th =0 indicates that no Gaussian filter was applied).
Good synthetic LCs have been obtained with between 3 and 9 delay values which,
depending on the amount of smoothing, range from $\tau = 15$ to 27\th d.  With
larger smoothing, ($\sigma_s >0.03$), the reconstructions deteriorate, with the
synthetic LCs becoming more symmetric about the average and more regular.
Figure~\ref{rumisyn} displays a sample of our best segments of 4D synthetic
LCs, with the delays $\tau$, and the smoothings $\sigma_s$ and $W_g$ indicated
on the right.  The fact that our 4D reconstructions appear to capture the
dynamics of the pulsations of R~UMi suggests that $d_e=4$.

 \begin{figure*}
\centerline{\psfig{figure=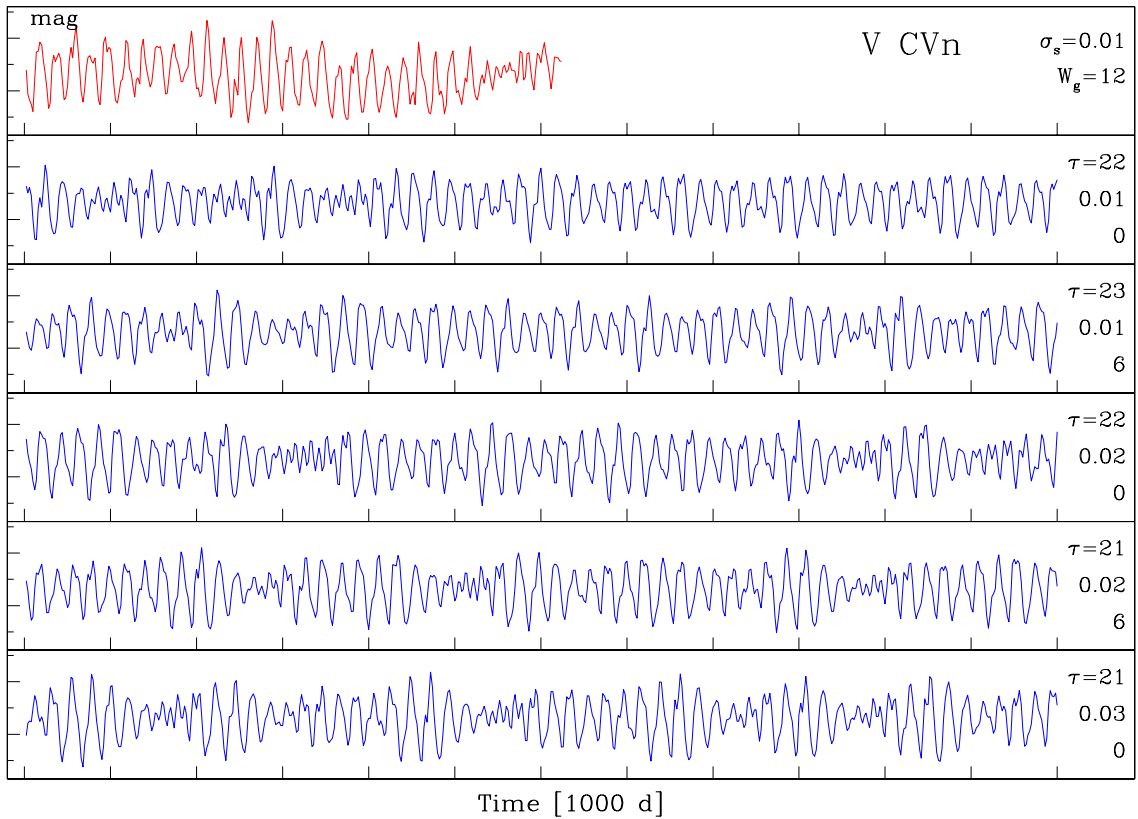,width=15.cm}}
\caption[]{\small   {\bf V CVn:}
top: smoothed LC (mag);
below:
sections of synthetic LCs with various seeds and delays
$\tau$ for
maps constructed from the smoothed magnitude data;
($\tau$,$\sigma_s$,$W_g$) are indicated on the right
$W_g=0$ stands for no Gaussian filter);
below:
sections of synthetic LCs with various seeds
and delays $\tau$;
maps constructed from the smoothed magnitude data;
   }
 \label{vcvnsyn}
 \vskip -5pt
 \end{figure*}

 \begin{figure*}
 \centerline{\psfig{figure=
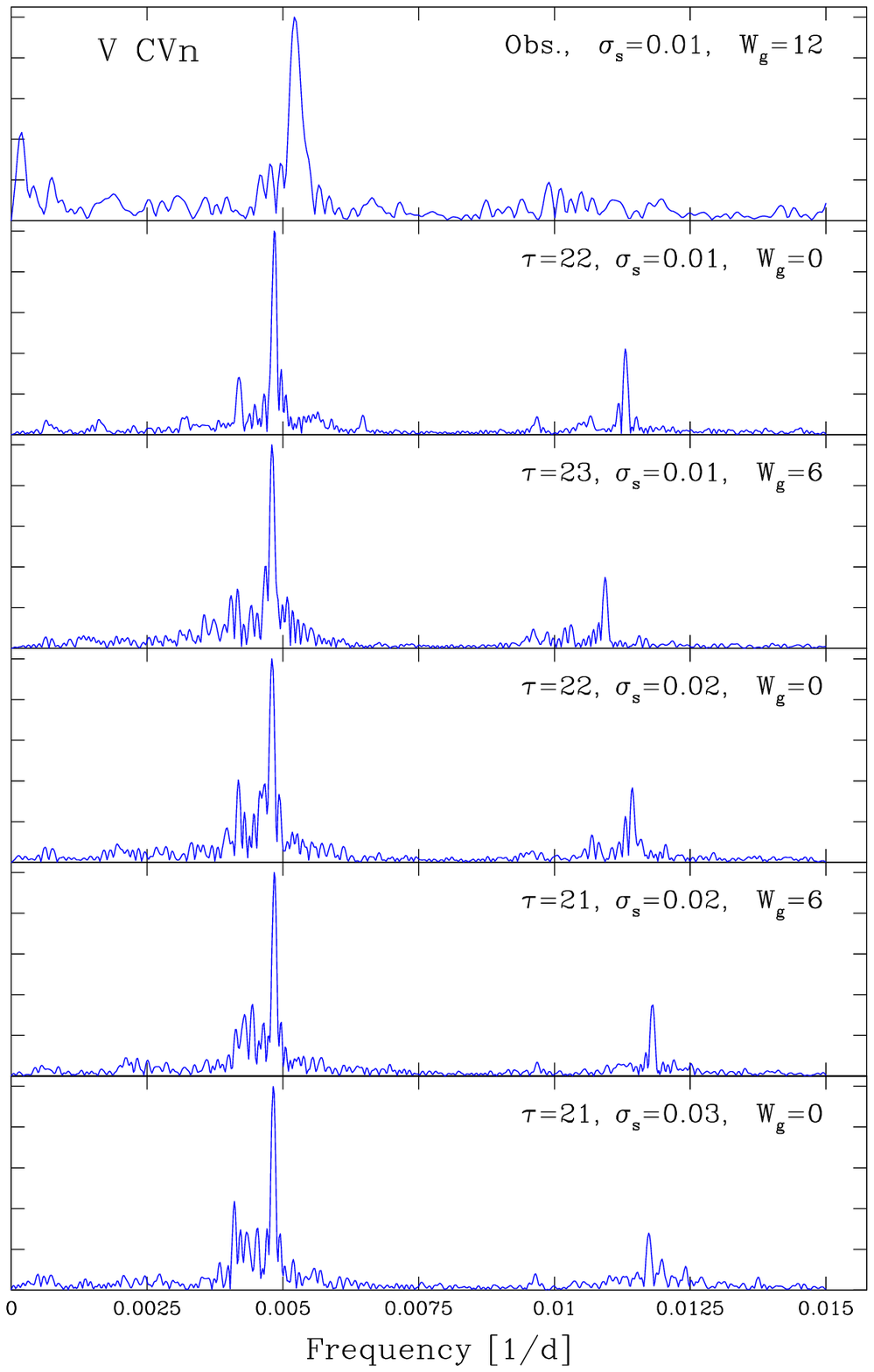
,width=8.1truecm}
             \psfig{figure=
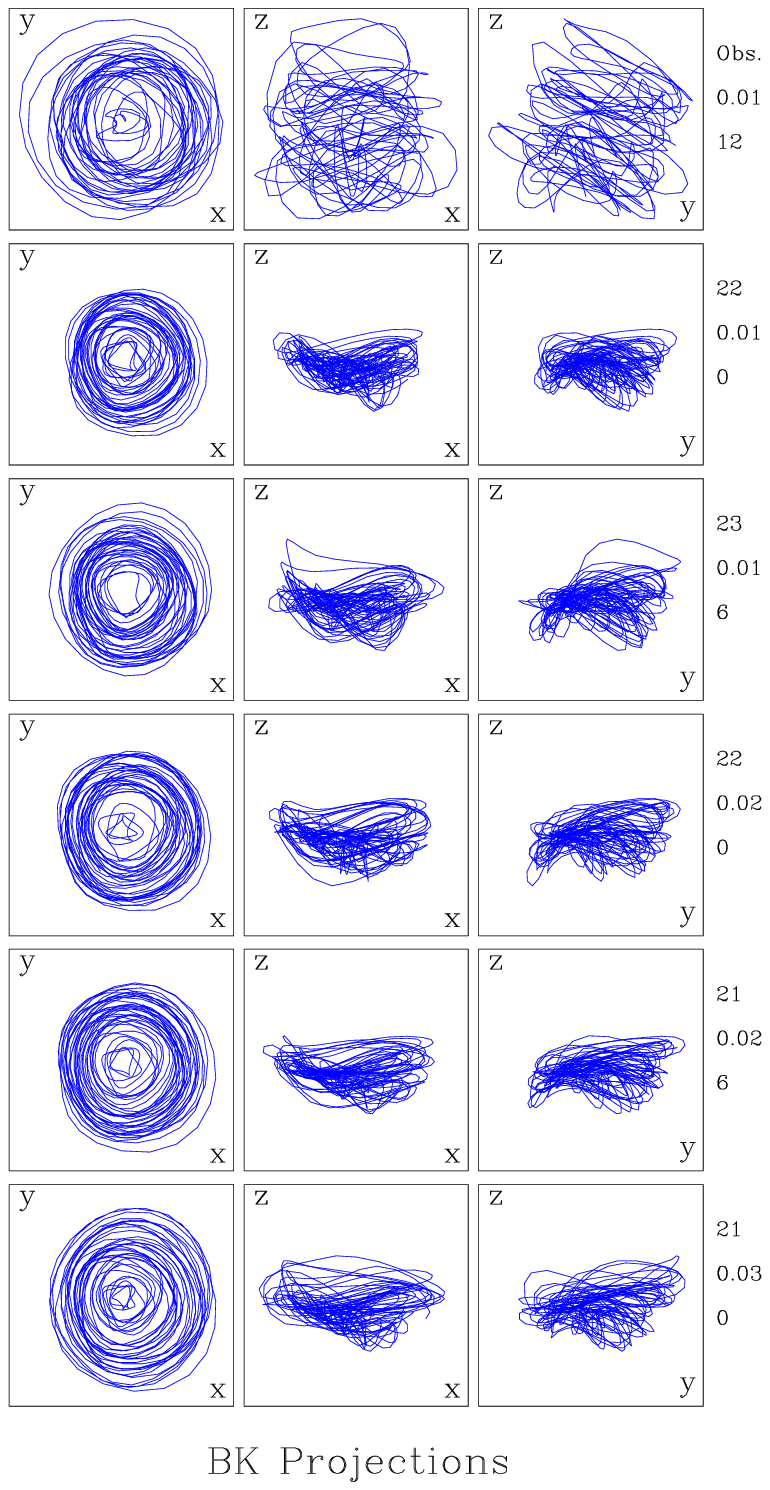
,width=6.60truecm} }
\caption[]{\small {\bf V~CVn}:
LEFT: Amplitude FS
top: of the smoothed LC,
below: of synthetic LCs.
 RIGHT:  corresponding Broomhead-King (BK) projections.
   }
 \label{rscygfs}
 \end{figure*}

A second comparison can be made at the level of the {\sl amplitude Fourier
spectra (FS)}.  Because the signals are unsteady, we do not want to compare
individual peaks, but rather the FSs should have the same overall structure.
Figure~\ref{rumifs} displays successively from top to bottom, the FS of the
smoothed and of our best synthetic LCs, indicating excellent agreement.  The
relative broadness of the fundamental peak for the observed LC is due to its
shorter time span.


 \begin{figure*}
 \centerline{\psfig{figure=
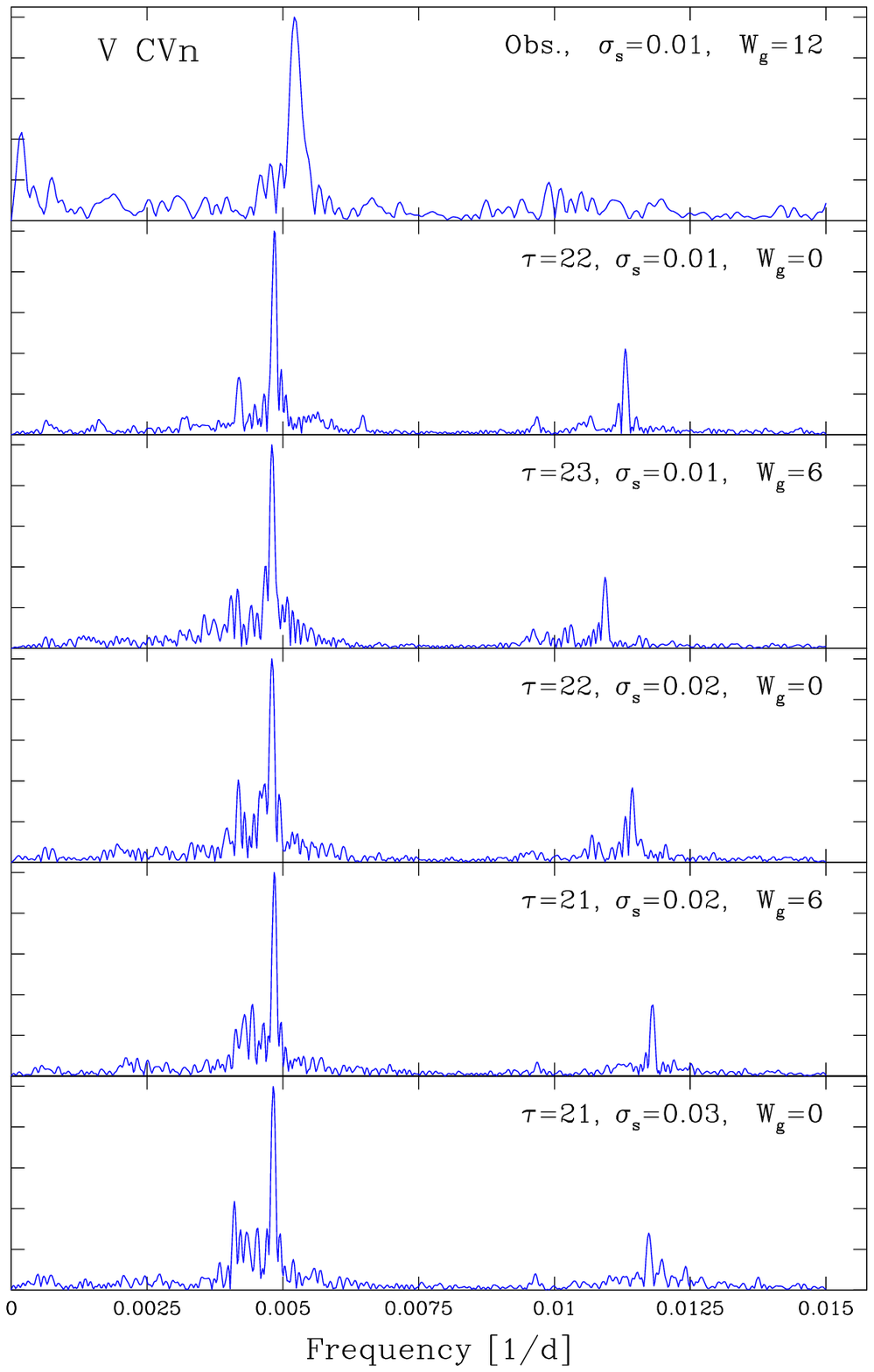,width=8.1truecm}
             \psfig{figure=
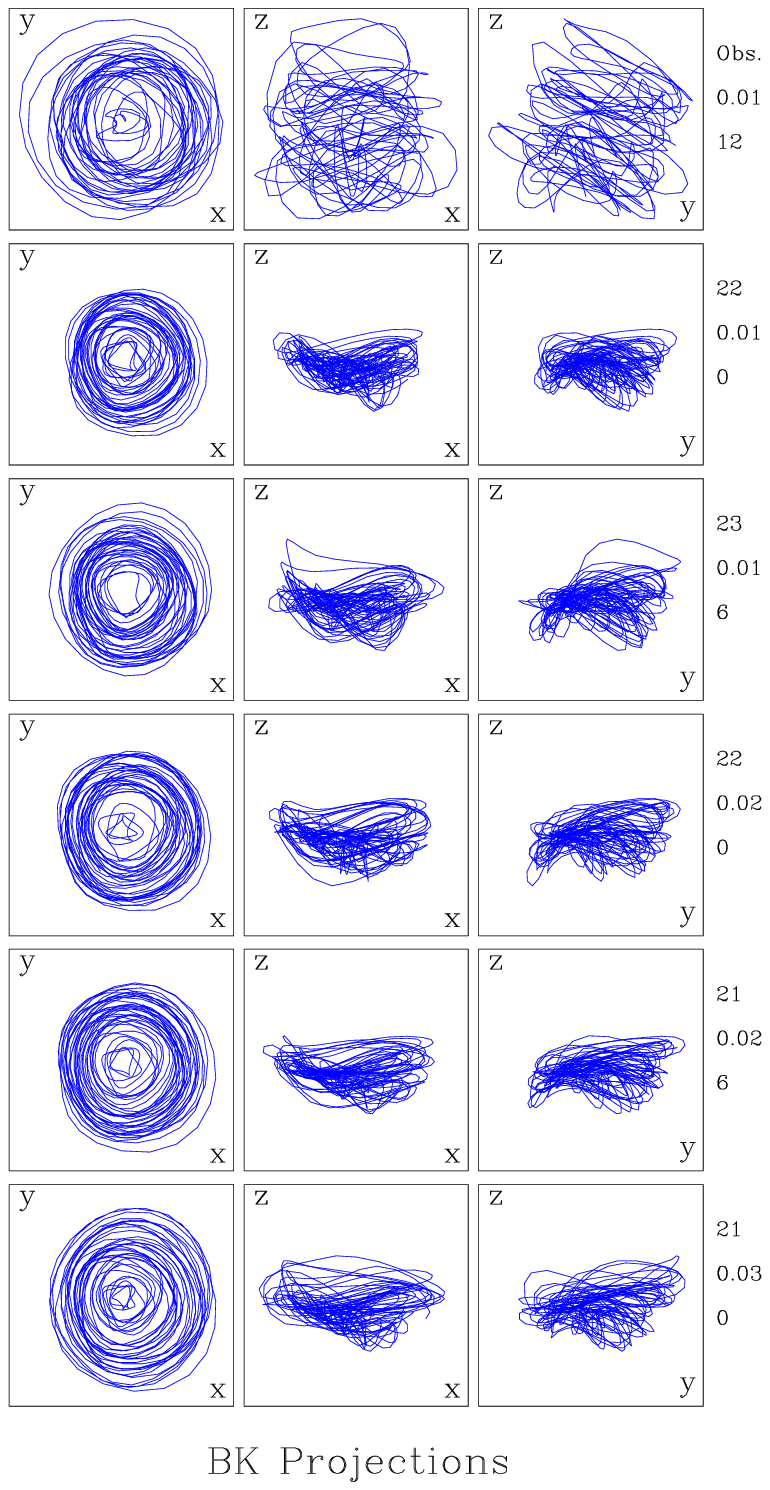,width=6.60truecm} }
 \caption[]
  {\small {\bf V CVn:}
   LEFT: Amplitude FS
   top: of the smoothed LC,
   below: of synthetic LCs.
 RIGHT:  corresponding Broomhead-King (BK) projections.
    }
 \label{vcvnfs}
 \end{figure*}

Finally, a third comparison can be performed with {\sl BK projections}
(Broomhead \& King 1987) which are the projections of the temporal signals on
the first few eigenvectors of the auto-correlation matrix which represent an
optimal visual way of spreading out the attractor.  In practice, we first
generate the projection matrix with the smoothed observed LC with a chosen
delay $\tau$, and then use this matrix to make the projections of all the LCs.
The BK projections onto the first three eigenvectors, labelled x, y and z, are
displayed in Fig.~\ref{rumifs}, indicating that the structure of the
reconstructed attractor is very similar to that of the observations.

The three tests indicate that, considering the extreme shortness of the data,
the reconstructions for R~UMi are quite good and reasonably robust.

In Table~\ref{tab} we present {\bf the linear stability eigenvalues} ($i 2 \pi
f + \kappa$) of the fixed points of the maps.  They should correspond to the
frequencies of the modes that are excited in the pulsation and therefore to the
stability of the equilibrium state of the star.  Note that the value of the
lowest frequency $f_1$ is close to the frequency of the lowest peak in the FS
(0.00333$\pm$0.00003\th d$^{-1}$).  The frequency $f_2$ is not as well
determined, although the frequency ratios are reasonably close to a 2:1
resonance.  

Table~\ref{tab} indicates that the growth rates are comparable to the
frequencies, indicating the presence of extreme nonadiabaticy, consistent with
the numerical modelling of such stars (Kov\'acs \& Buchler, 1988).  Column 4
represents the values of the magnitude of the fixed point, \ie the 'center of
the motion', which are close to the average magnitude of the observed LC, \viz
0.95, but not equal to it presumably because of the skewedness of the light
curve.  The growth rate $\kappa_1$ of the lower frequency mode is positive
(unstable) whereas the second mode is damped ($\kappa_2<0$).  We see that the
reconstructions consistently reproduce the linear part of the maps, despite the
fact that the amplitude of the LC never gets very small, and that
therefore it does not explore the neighborhood of the fixed point.

We note in passing that in nonlinear dynamics this scenario of two resonant
modes, one unstable and the other stable, goes under the name of modified
Shil'nikov scenario (Glendenning \& Tresser 1985).

Lyapunov exponents and the (generally non-integer) {\sl Lyapunov dimension
$d_L$} (\eg Ott 1993, Serre \etal 1996) can give us additional quantitative
information about the attractor.  In particular $d_L$ sets a lower limit on the
dimension $d$ of the physical phase space, because an attractor of fractal
dimension $d_L$ must live in a Euclidean space of dimension $d > d_L$.  For
example, a dynamical system of two coupled pulsation modes ($d=2\times 2 =4$)
can have an attractor with a fractal dimension $d_L \leq 4$.

In order to extract accurate Lyapunov exponents and a Lyapunov dimension $d_L$
one needs LCs containing at least 500 cycles, \ie $\sim$ 200,000\th d.
The observed LC with 6500\th d is much too short for that purpose.  Furthermore
the maps for R~UMi are not stable enough to generate sufficiently long
synthetic LCs to compute reliable Lyapunov exponents, but they yield a rough
estimate.  For 12000\th d long synthetic LCs with $\sigma_s=0.01 - 0.03$ the $d_L$
fall in a broad range from 2. to 3.2 with an average of 2.8.  For the best few
synthetic LCs $d_L$ averages to 2.9.  A couple of rather rare, 130,000\th d
long 4D iterations, with $\sigma_s=0.01$, $W_g=8$, $\tau=23$ gave $d_L=2.9$
and for $\sigma_s=0.01$, $W_g=8$, $\tau=19$ gave $d_L=3.1$.  We believe that
these latter values may be closer to the actual value of $d_L$, but these
values should be taken with a grain of salt.

In principle, 5D reconstructions can be used to confirm that a 4D space is
large enough, because the fractal dimension should be independent of the
dimension of the reconstruction space.  Unfortunately, such reconstructions are
not very robust because of the paucity of pulsation cycles in the observed LC.
We note however that the $d_L$ for the few decent 5D synthetic LCs that we
could make cluster around 3.5, \ie they essentially do not increase much as
$d_R$ is increased from 4 to 5.  This is a corroboration that the embedding
dimension is indeed $d_e=4$.

In the study of R~Sct (Buchler \etal 1995, 1996), with the help of $d \geq d_L
\sim 3.1$, we were able to obtain a {\it lower limit} of 4 for the dimension
$d$ of the physical phase space of the dynamics, and very strong evidence for a
minimum embedding dimension $d_e=4$ gave an {\it upper limit} of $d=4$.  We
thus could infer that $d=4$ for R~Sct.  Here, for R~UMi we can state the
minimum $d_e=4$ and thus the upper limit on $d=$ with only some confidence.
Also, because of the uncertainties in $d_L$ just mentioned we obtain a lower
limit of either 3 or 4 for $d$, depending whether $d_L < 3$ or $d_L> 3$.
However, if $d$ were only equal to 3, then one would have to invoke the
coupling of one oscillatory mode (2D) with a thermal (secular) mode (1D),
rather than with a second oscillatory pulsation mode (2D).  While this is
possible we deem it unlikely, and it appears that $d=4$, as for R~Sct.  In this
4D physical phase space, the complex amplitudes of the two pulsation modes may
be considered the natural generalized coordinates for this dynamics.

Furthermore, the linear stability analysis of the 4D maps gives us results that
are very similar to those of R~Sct (Buchler \etal 1996).  We can therefore
conclude that the same pulsation mechanism is operative in R~UMi, namely that
its irregular LC is generated by two nonlinearly interacting, resonant
modes in which the low frequency mode is linearly unstable and the harmonic
mode is linearly stable, resulting in a continual and unsteady energy exchange.

 \begin{figure*}
\centerline{\psfig{figure=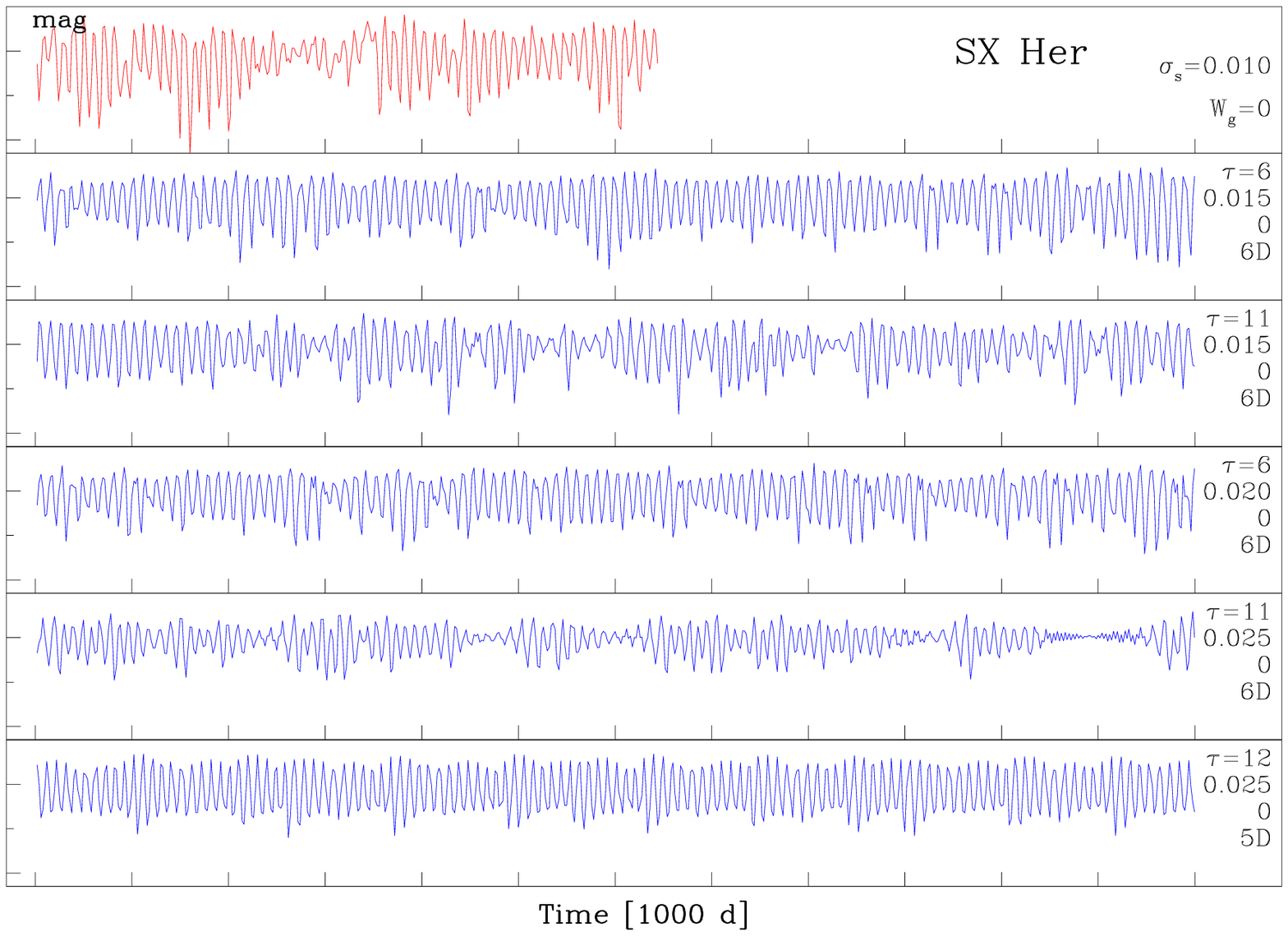,width=15.cm}}
\caption[]{\small  {\bf SX~Her:}
top: smoothed LC (mag);
below:
sections of synthetic LCs with various seeds and delays
$\tau$ for maps constructed from the smoothed magnitude data;
($\tau$,$\sigma_s$,$W_g$) are indicated on the right
($W_g=0$ stands for no Gaussian filter);
   }
 \label{sxhersyn}
 \end{figure*}


 \begin{figure*}
 \centerline{\psfig{figure=
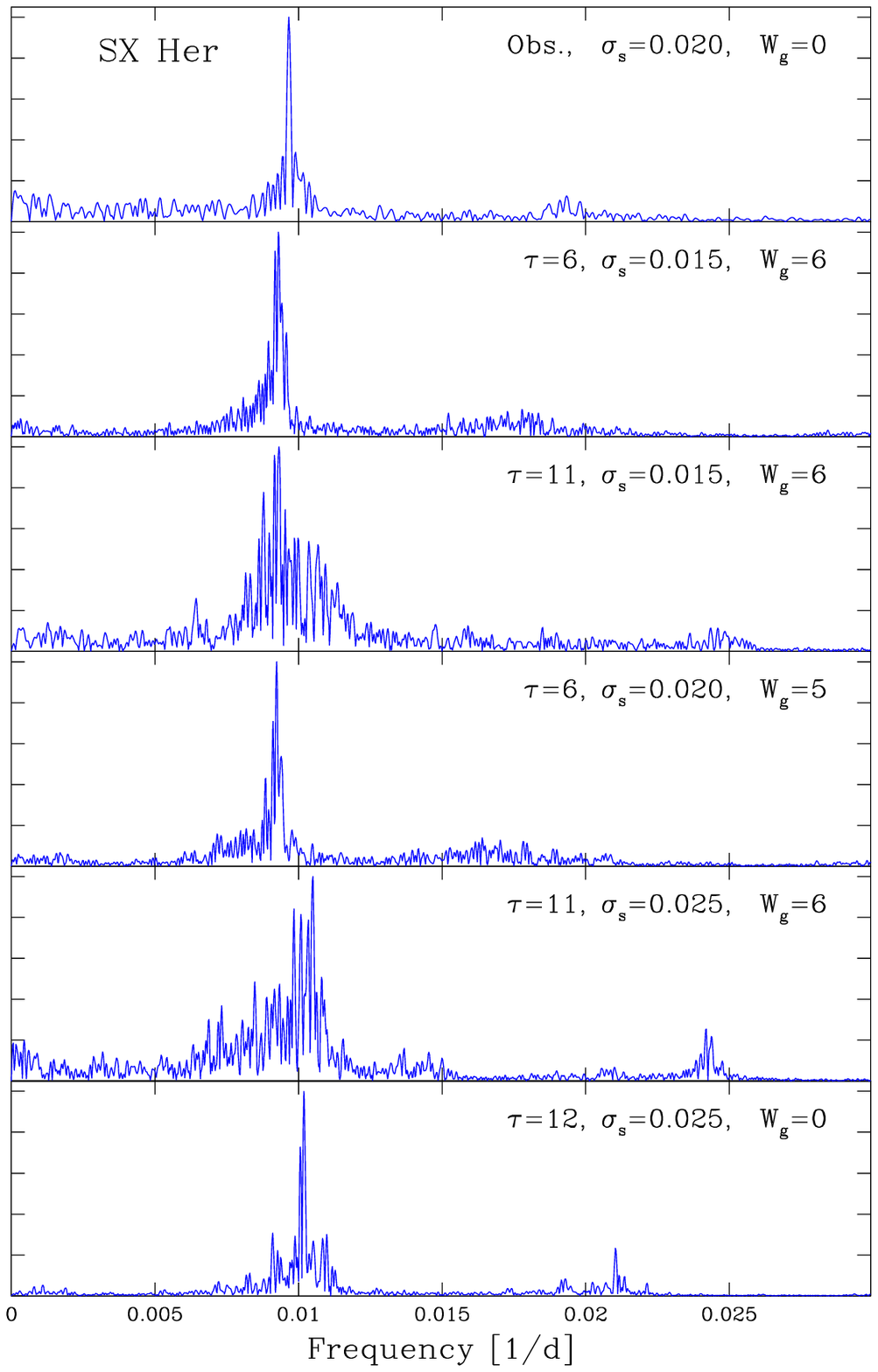
,width=8.1truecm}
             \psfig{figure=
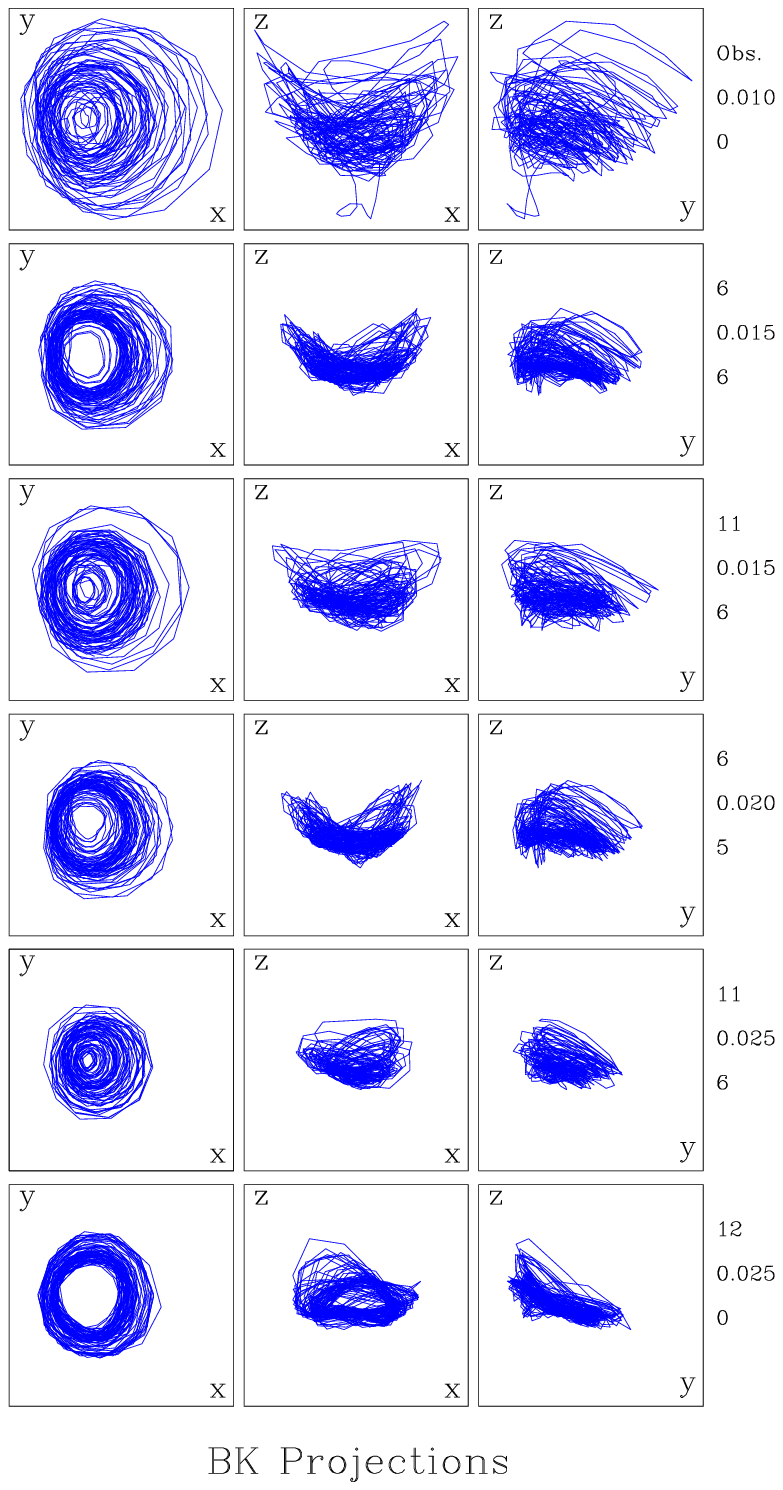
,width=6.63truecm} }
 {\small {\bf SX~Her:} Fourier amplitude spectrum
 top: of the smoothed LC,
 below: of synthetic LCs.
 BK projections,
  top: smoothed LC,
  below:
  sections of synthetic LCs with various smoothings
  and delays $\tau$.   }
 \label{sxherfs}
 \end{figure*}


\vskip -1.5cm
\subsection{\bf RS~Cyg}

In Figure~\ref{rscygsyn} on top we display the 15 cycles long LC for RS~Cyg,
superposed on a LC fitted with the smoothings $\sigma_s=0.02$,
$W_g=4$.

We find again that 3D reconstructions are not satisfactory at all, with the
periods, in particular, being quite bad.  The 4D reconstructions show some
insensitivity to the smoothing, as long as it is large enough: Good LCs have
been obtained from a range of typically 2 -- 5 delay values which, depending on
the amount of smoothing, range from $\tau = 34$ to 40\th d.  Synthetic LC
obtained from reconstructions with a smoothing of $\sigma_s=0.010$ and 0.020
were too regular, and the fixed points of the map are about 0.20 to 0.25 too
large.  The displayed synthetic LCs come from smoothing values $\sigma_s =
0.025 - 0.04$.  Typically there are from 2 to 6 $\tau$ values for which there
are good synthetic LCs.  Fig.~\ref{rscygsyn} displays a sample of our best
12000~d long segments of 4D.  This suggests that the embedding dimension for
RS~Cyg is also $d_e=4$.  This conclusion is corroborated by the amplitude FS of
Fig.~\ref{rscygfs} and the BK projections in Fig.~\ref{rscygfs}.

The linear stability eigenvalues of the fixed points of the maps are shown in
Table~2.  The frequency ratios are approximately to 2:1 for the reconstructions
with the larger values of $\sigma_s \geq 0.30$ that we have used.  The growth
rate $\kappa_1$ of the first mode is again positive (unstable) and that of the
second mode $\kappa_2<0$, but not for $\sigma_s =0.025$.  In all cases, the
lowest frequency $f_1$ is reasonably close to the frequency of the fundamental
peak in Fig.~\ref{rscygfs}.  That the linear stability eigenvalues of the fixed
points are not very consistent is not very astonishing, because after all the
amplitude of the LC is never small, and therefore it does not at all
explore the neighborhood of the fixed point where the stability properties are
determined.

The Lyapunov dimensions $d_L$ of all our good synthetic LCs are found to run
from 2.1 to 3.2 with an average of $\sim 2.45$.  The same average $d_L$ is
obtained for the best few.  This imposes a lower limit of 3 or 4 on $d$, but
the latter can still be, and we believe it to be 4 for the same reasons we have
given in the discussion on R~UMi above.  There are unfortunately too few data
points to attempt a reconstruction in 5D which could corroborate that $d_e=4$.

We conclude that the analysis of the RS~Cyg LC is consistent with, and perhaps
suggests again that two interacting, highly nonadiabatic modes are at work,
which in addition are in resonance.  In other words RS~Cyg seems to have the
same underlying 4D dynamics that we have found for R~Sct and R~UMi.


\vskip -1.5cm
\subsection{\bf V~CVn}

The smoothed V~CVn LC is displayed in Fig.~\ref{vcvnsyn} ($\sigma_s=0.01$,
$W_g=12$).

We have not been able to make any satisfactory 3D reconstructions for V~CVn.
However, the 4D reconstructions show insensitivity to the smoothing.  For each
smoothing set, good synthetic LCs have been obtained for typically 2 -- 8 delay
values which, depending on the amount of smoothing, range from $\tau =17$ to
27\th d.  Based on the visual appearance of the synthetic LCs the 4D
reconstructions in Fig.~\ref{vcvnsyn} are good, suggesting $d_e=4$.  However,
for some reason, the dominant frequencies of all the synthetic LCs are
systematically smaller \th (Fig.~\ref{vcvnfs}), and the overtone frequencies
are somewhat larger.

 \begin{figure*}
\centerline{\psfig{figure=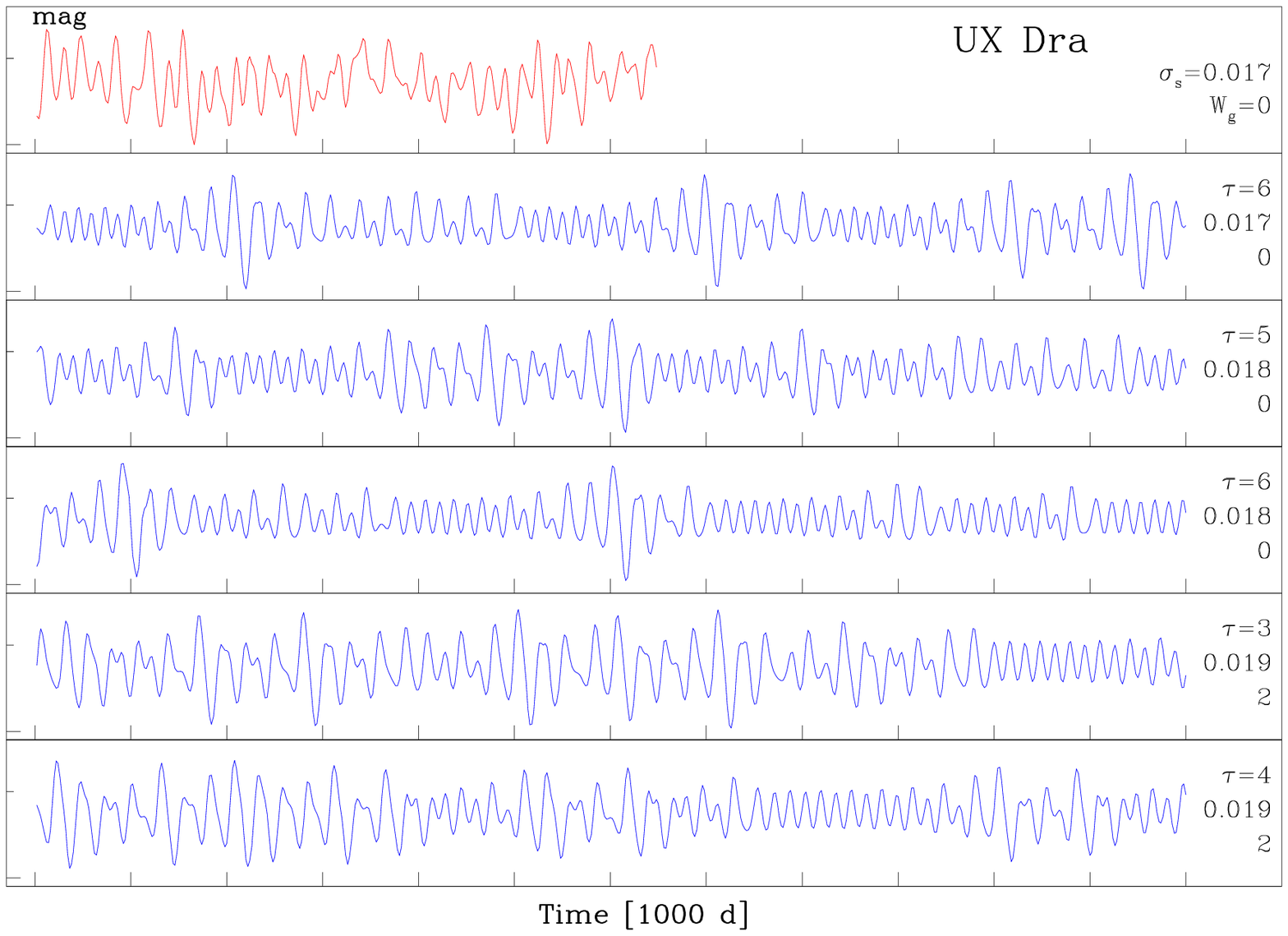,width=15.cm}}
\caption[]
  {\small  {\bf UX Dra:}
  top: smoothed LC (mag);
  below:
  sections of synthetic LCs with various seeds and delays
  $\tau$ for
  maps constructed from the smoothed magnitude data;
  ($\tau$,$\sigma_s$,$W_g$) are indicated on the right
  ($W_g=0$ stands for no Gaussian filter);
   }
 \label{uxdrasyn}
 \end{figure*}


 \begin{figure*}
 \centerline{\psfig{figure=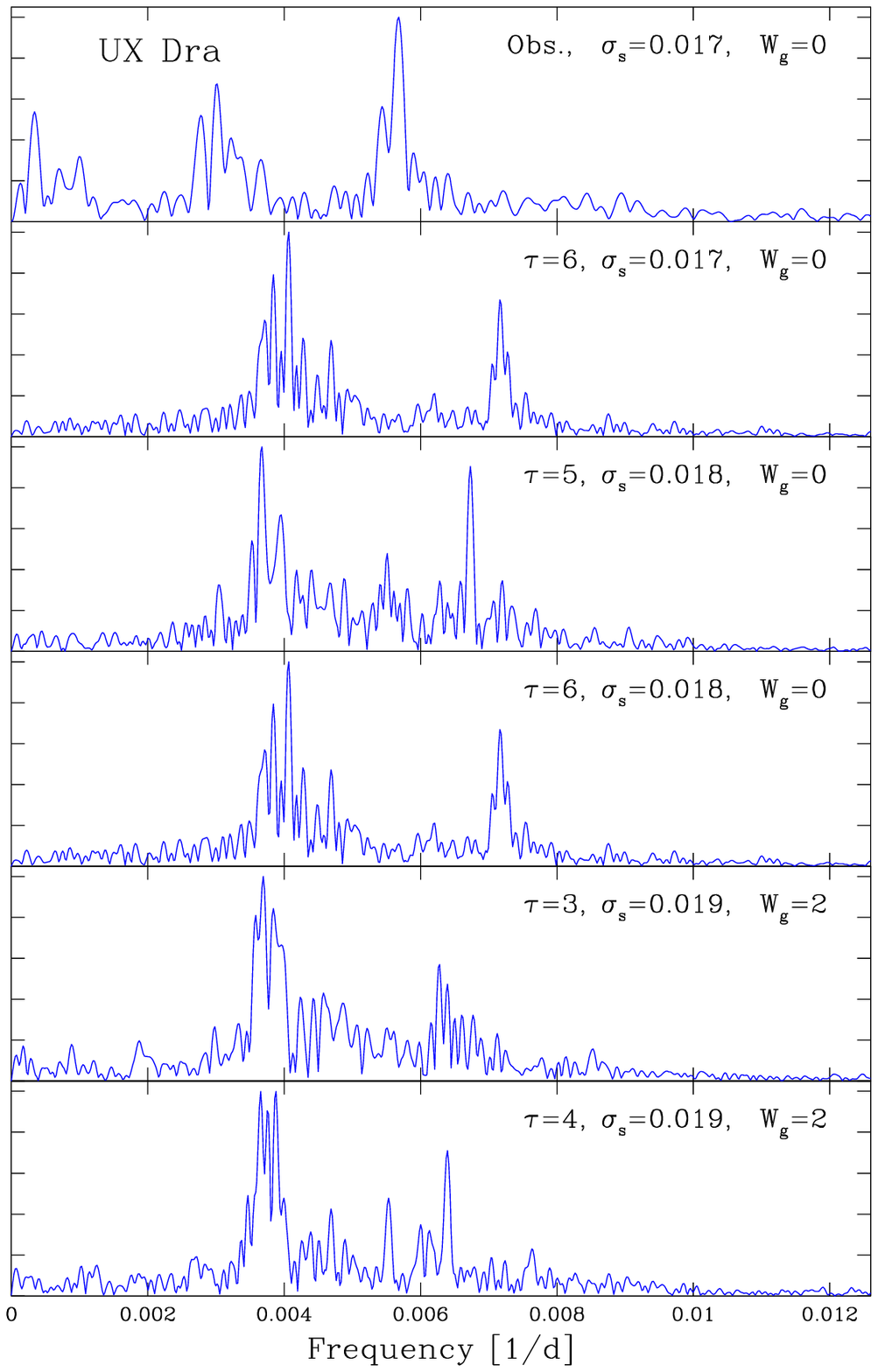,width=8.1truecm}
             \psfig{figure=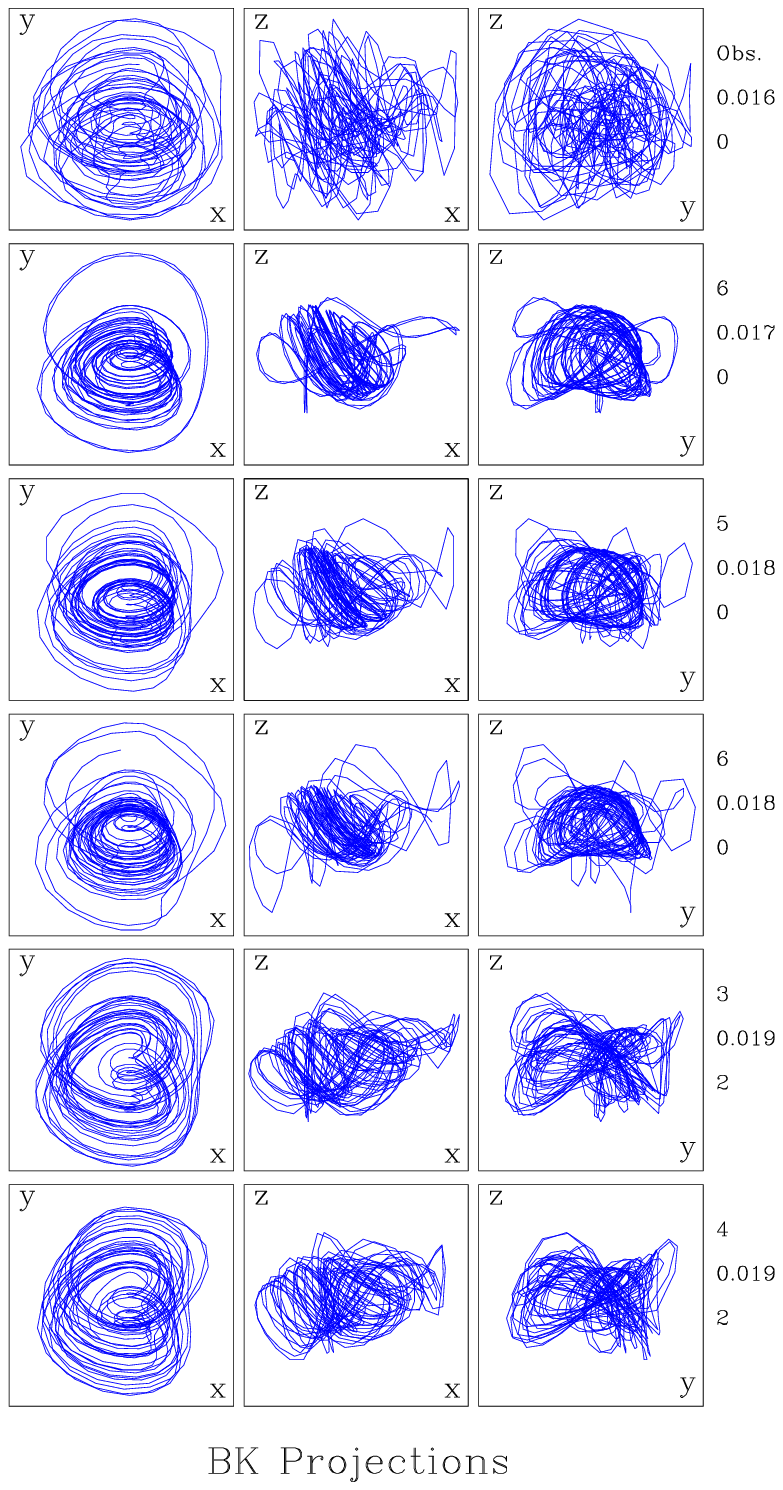 ,width=6.63truecm} }
 \caption[]
  {\small {\bf UX Dra:} amplitude FS,
  top: of the smoothed LC,
  below: of synthetic LCs.
 BK projections,
  top: of the smoothed LC,
  below: of synthetic LCs.
   }
 \label{uxdrafs}
 \end{figure*}

The BK projections, displayed in Fig.~\ref{vcvnfs}, are a little tamer for the
synthetic LCs.  The same difference is already apparent in Fig.~\ref{vcvnsyn},
where the LCs show fluctuations in the average magnitude that are stronger than
those of any of the synthetic ones.  It could be that these fluctuations of the
average magnitude are due to dust, or it could be that the dynamics is more
complicated in the sense that a higher dimensional phase space (more modes) is
required to capture it.  In order to answer that question we would likely need
a longer LC.

The Lyapunov dimensions $d_L$ of the good 4D synthetic LCs range from 3.4 to
3.6, substantially larger than for R~UMi and RS~Cyg.  Good 5D reconstructions
are also possible, but they occur only for one or 2 $\tau$ values that run from
15 to 26\th d, depending on the smoothing.  Significantly, the fractal
dimensions of the good synthetic LCs run from $d_L=3.65$ to 4.02, \ie they do
not increase with $d_R$, corroborating that indeed $d_e=4$.

Another interesting difference with R~Sct or R~UMi appears in the linear
stability eigenvalues of the fixed point displayed in Table~\ref{tab}.  The
frequencies associated with the linear stability eigenvalues of the fixed point
are again in a close resonance ratio of 2:1, and the modes are highly
nonadiabatic.  However, here it is the higher frequency mode that is unstable,
and the lower one that is stable.

In the study of Buchler \etal 2001 reconstructions were unsuccessful because
the range of smoothing parameters was very limited.  This shows that a survey
with a range of smoothings is important.


\vskip -1.5cm
\subsection{\bf SX~Her}

The smoothed LC of SX~Her is shown on top of Fig.~\ref{sxhersyn} with
($\sigma_s=0.01$, $W_g=0$).  Underneath, are our best segments of synthetic LCs
in several reconstruction dimensions $d_R$.

As for the previous stars, for SX~Her the 3D reconstructions are not good at
all.  In 4D they are a little better, but still not very satisfactory, having
more symmetry than the LC.  As already found in Buchler \etal (2002) they are
very sparse in $\tau$ space, and hence not robust.  5D reconstructions, while
capturing some of the features of the LCs appear unsatisfactory.  On
the other hand, in 6D the synthetic LCs display many of the characteristic
features of the LC, such as asymmetric bursts.

The Fourier spectra are shown in Fig.~\ref{sxherfs} for the same 
LCs displayed in Fig.~\ref{sxhersyn}.  The first frequency is
recovered relatively well, although the fundamental peak shows more structure
than for the LC.  The BK projections shown in Fig.~\ref{sxherfs} are
reasonably good, but they reflect the tamer nature of the LCs and
their smaller amplitudes.

We conclude that the dynamics which underlies the pulsations of SX~Her may be
more complicated than that of the other stars (to witness, for example, the
intermittency in the central part of the analyzed LC) and therefore
higher dimensional ($d>4$, perhaps $d=6$).  However, this conclusion should be
taken with some caution because the observational coverage of the LC
is sparse in some places for this shorter 'period' star, as Fig.~\ref{cad}
shows.  It is also possible that we have missed a range of smoothing parameters
where the lower-D reconstructions are successful.

\vskip -1.5cm
\subsection {\bf UX~Dra}

Figure~\ref{uxdrasyn} on top displays a smoothed fit to the LC for UX~Dra \th
($\sigma_s=0.017$, $W_g=0$).

In a 3D reconstruction space the synthetic LCs definitely do not look like the
LC.  In 4D good looking synthetic LCs can be generated, although the
reconstructions are not particularly robust.  They appear possible only for a
relatively narrow range of $\sigma_s$ and only for a few $\tau$ values in the
range 3 -- 6\th d.  This relative lack of robustness is not very astonishing
considering that, compared to its complexity, the LC is very short.
For a good reconstruction we clearly need LC features to repeat over
the span of the LC.  In addition the amplitude of this star is only half that of
the other 4 stars, so that the noise level is higher.  For the same reasons we
have had difficulty making reconstructions in 5D and 6D.  However, the
synthetic LCs are consistent with $d_e=4$.

Figure~\ref{uxdrasyn} displays a sample of our best segments of 4D ynthetic
LCs.  The dominant frequencies of all the synthetic LCs are not very good as
seen in the amplitude FS in Fig.~\ref{uxdrafs}.  The BK projections of the
synthetic LCs, although they bear a good resemblance to that of the LC, but are
tamer.  The Lyapunov dimensions for the synthetic LCs of Fig.~\ref{uxdrasyn}
range from 2.4 to 3.1.

The linear stability analysis, shown in the Table~\ref{tab}, as for V~CVn
yields a {\sl stable} low frequency, $f_1$, mode and an unstable $f_2$ mode,
the stability being the reverse of that of R~Sct and R~UMi, but the same as
for V~CVn.  However, the $f_1$ frequencies are not particularly close to the low
frequency peak $\sim 0.003$\th d$^{-1}$ in the FS, and the frequency ratio has
a larger scatter around 2:1 than for the other stars.

We are forced to conclude that, either the observed LC of UX~Dra are too short
to make robust reconstructions, or perhaps the dynamics could be contaminated
by dust formation and obscuration.

\vskip -1.5cm
\subsection{Discussion}

In our survey we have only varied the dimension $d_R$, the smoothing parameters
$\sigma_s$ and $W_g$ and the delays $\tau$.  The polynomial order has been
fixed at $p=4$.  For the binning and averaging time interval we have chosen
$\Delta t=10$d for all stars except UX~Dra, where we obtained better results
with $\Delta t=1$d.  In general, we have not attempted to optimize the values
of $\Delta t$ which might lead to more robust reconstructions.

Each vector component of our polynomial maps in 4D contains 70 coefficients,
and one might be inclined object that it is not astonishing that such a map
fits the data well, and that one would obtain an equally good fit with a
(linear) multi-periodic Fourier sum with the same number of coefficients.
However, for the latter is just a fit with limited predictive power and which
tells us very little about the actual dynamics.  In contrast, the constructed
(nonlinear) maps have remarkable predictive power, \ie the long synthetic LCs
that we generate from the map do capture the dynamics as our comparisons have
shown.

\vskip -1.5cm
\subsection{Comparison with the classical variable stars}

The reader may be curious why in contrast to the Cepheids, which are the close
relatives of the semiregular variables, pulsate in a regular, \ie periodic or
multiperiodic fashion.  The fundamental physical difference is in the size of
the relative linear growth rates (\ie the growth rates $\times$ periods).  For
the classical variable stars, the latter are very small, of the order of a
percent, whereas those of the Population II Cepheids, semiregulars and Mira
variables are of order unity.  The reason for this difference lies in the fact
that the luminosity/mass ratio of the latter is about ten times larger.  This
causes a large increase in the coupling between pulsation and heat flow and
concomitantly in the growth rates.

When the relative growth rates of the excited modes are small, as in the
classical Cepheids, there exists near center manifold in the parlance of
nonlinear dynamics, \eg Buchler 1993).  It can then be shown that the most
complicated motions will be periodic pulsations (limit cycles) or multiperiodic
pulsations (\eg beat Cepheids).

No such manifold exists in the highly nonadiabatic semiregular stars, and more
complicated, irregular (chaotic) pulsations can arise.  Intuitively, large
growth rates are a prerequisite for irregular pulsations, \ie for amplitude
modulation on the time scale of a period, although this is not sufficient.

It was shown unambiguously that the Hertzsprung progression of the Fourier
phases of the bump Cepheids has its origin in a 2:1 resonance between the
linearly unstable fundamental mode and the linearly stable second overtone
(Buchler 1993).  It is again the presence of a center manifold that gives rise
to a synchronization of the two modes, and a periodic pulsation results in
which the presence of the entrained second mode merely gives rise to a bump on
the light curve.  In contrast, although the same 2:1 resonance occurs in the
Semiregulars, the absence of a center manifold allows the pulsation to be
chaotic.

\section{CONCLUSIONS}

We have analyzed the 6500\th day long observed LCs of the semiregular/Mira
stars R~UMi, V~CVn, RS~Cyg, SX~Her and UX~Dra.  First, with a time-frequency
analysis we have shown that in each case the dominant instantaneous frequency
peak is not steady, and that the 'harmonic peak' does not vary synchronously
with the dominant one.  This has led to the conclusion that these stars are not
multi-periodic in the usual sense of the word, but that they are more likely to
be generated by a low dimensional chaotic dynamics that involves a small number
of pulsations modes.  {\it Multimode} might therefore be a more suitable label
for these stars than multiperiodic.

While a purely mathematical analysis might lead one to a stochastic nature of
these pulsations, we have clearly ruled this out on physical grounds.

The results of a global flow reconstruction technique give strong evidence that
the LCs of all five stars are generated by a low dimensional chaotic pulsation
dynamics.  The observed LCs of all 5 stars are really very short in terms of
the number of pulsation cycles and their complexity, and our results must be
taken with a little caution.  However, interestingly, the evidence is very
strong that for R~UMi, RS~Cyg and V~CVn, and probably for UX~Dra as well, the
pulsation takes place in a 4D phase space.  The LC of SX~Her appears to be more
complicated and perhaps a higher dimensional space, 6D, may be required.

Our analysis shows that the LCs of R~Umi and RSC~Cyg are generated by nonlinear
dynamics that involve two resonant pulsation modes, a linearly self-excited
lower frequency mode which entrains a second, linearly stable mode through a
2:1 resonance.  The irregular pulsation occurs as a result of continual
exchange of energy between the two resonant modes.  This is the same scenario
as was found for the irregularly pulsating RV~Tau-type star R~Sct (Buchler
\etal 1995, 1996), but we caution that the reconstructions are not as robust as
they are for R~Sct because of the shortness of the observed light curves.  In
particular, we were not able to construct long synthetic LCs and extract a
reliable Lyapunov dimension from these as for R~Sct.  The analysis of RS~Cyg
strongly suggests the presence of the same pulsation mechanism is operative.

Interestingly, the analysis shows that the LC of V~CVn may similarly be
due to a resonant interaction of two pulsation modes in a 2:1 resonance, but
here it appears that the higher frequency mode is self-excited, whereas the
lower frequency mode is stable.  The analysis of UX~Dra is much less
conclusive, but is compatible with the same resonant 2-mode scenario.

To summarize then, we find good evidence for a common physical explanation for
the irregularity that characterizes the pulsations of R~UMi, RS~Cyg, V~CVn (and
probably UX~Dra) as we had found earlier for R~Sct, namely that the underlying low
dimensional chaotic pulsation dynamics arises from the nonlinear interaction of
two resonant modes, and their high nonadiabaticity causes the pulsations to be
chaotic.

\acknowledgements

This work has been supported by the National Science Foundation (AST03-07281)
at UF and (AST-8420901 and AST-8718533) at Grinnell College, and by the
Hungarian OTKA (T-038440, T-038437).  We thank the AAVSO, AFOEV, BAAVS and
VSOLJ for the data that were used in Fig.~\ref{vcvnzk}.  This research has made
use of the SIMBAD database, operated at CDS, Strasbourg, France. We also wish
to thank an anonymous referee for his comments.

\end{document}